\documentclass[useAMS,usenatbib]{mn2e}
\usepackage[dvips]{graphicx}
\usepackage[T1]{fontenc}

\def\lesssim{\mathrel{\hbox{\rlap{\hbox{\lower4pt\hbox{$\sim$}}}\hbox{$<$}}}}
\def\gtrsim{\mathrel{\hbox{\rlap{\hbox{\lower4pt\hbox{$\sim$}}}\hbox{$>$}}}}
\newcommand{\mincir}{\raise-2.truept\hbox{\rlap{\hbox{$\sim$}}\raise5.truept
\hbox{$<$}\ }}
\newcommand{\magcir}{\raise-2.truept\hbox{\rlap{\hbox{$\sim$}}\raise5.truept
\hbox{$>$}\ }}
\newcommand{\be}{\begin{equation}}
\newcommand{\ee}{\end{equation}}
\newcommand{\ba}{\begin{eqnarray}}
\newcommand{\ea}{\end{eqnarray}}

\title[ShaSS: Survey overview and galaxy density]{Shapley Supercluster
  Survey (ShaSS): Galaxy Evolution from Filaments to Cluster Cores}

\author[Merluzzi et
al.]{P. Merluzzi$^{1}$\thanks{merluzzi@na.astro.it},
  G. Busarello$^{1}$, C.~P. Haines$^{2}$, A. Mercurio$^{1}$,
  N. Okabe$^{3}$, K.~J. Pimbblet$^{4,5}$, \newauthor
  M. A. Dopita$^{6,7}$, A. Grado$^{1}$, L. Limatola$^{1}$,
  H. Bourdin$^{8}$, P. Mazzotta $^{8}$, \newauthor
  M. Capaccioli$^{9}$, N.~R. Napolitano$^{1}$, P. Schipani$^{1}$ \\ merluzzi@na.astro.it \\
  $^1$ INAF-Osservatorio Astronomico di Capodimonte, Via Moiariello 16
  I-80131 Napoli, Italy\\ $^2$ Departamento de Astronom\'{i}a,
  Universidad de Chile, Casilla 36-D, Correo Central, Santiago, Chile
  \\ $^3$ Kavli Institute for the Physics and Mathematics of the
  Universe (WPI), Todai Institutes for Advanced Study,University of
  Tokyo, 5-1-5 Kashiwanoha, Kashiwa, Chiba 277-8583, Japan. \\ $^4$
  Department of Physics and Mathematics, University of Hull,
  Cottingham Road, Kingston-upon-Hull, HU6 7RX, UK \\ $^5$ School of
  Physics, Monash University, Clayton, Melbourne, Victoria 3800,
  Australia \\ $^6$ Research School of Astronomy and Astrophysics,
  Australian National University, Cotter Rd., Weston ACT 2611,
  Australia \\ $^7$ Astronomy Department, Faculty of Science, King
  Abdulaziz University, PO Box 80203, Jeddah, Saudi Arabia \\ $^8$
  Dipartimento di Fisica, Universit\`a di Roma
  Tor Vergata, Via della Ricerca Scientifica 1, I-00133 Roma, Italy \\
  $^9$ Dipartimento di Fisica, Universit\`a Federico II, Napoli,
  Italy}

\begin{document}

\date{Accepted . Received }

\pagerange{\pageref{firstpage}--\pageref{lastpage}} \pubyear{}

\maketitle

\label{firstpage}

\begin{abstract}
   We present an overview of a multi-wavelength survey of the Shapley
   supercluster (SSC; z$\sim$0.05) covering a contiguous area of
   260\,$h^{-2}_{70}$Mpc$^2$ including the supercluster core. The
   project main aim is to quantify the influence of cluster-scale mass
   assembly on galaxy evolution in one of the most massive structures
   in the local Universe. The Shapley supercluster survey (ShaSS)
   includes nine Abell clusters (A\,3552, A\,3554, A\,3556, A\,3558,
   A\,3559, A\,3560, A\,3562, AS\,0724, AS\,0726) and two poor
   clusters (SC\,1327- 312, SC\,1329-313) showing evidence of
   cluster-cluster interactions. Optical ($ugri$) and near-infrared
   ($K$) imaging acquired with VST and VISTA allow us to study the
   galaxy population down to m$^\star$+6 at the supercluster
   redshift. A dedicated spectroscopic survey with AAOmega on the
   Anglo-Australian Telescope provides a magnitude-limited sample of
   supercluster members with 80\% completeness at $\sim$m$^\star$+3.

  We derive the galaxy density across the whole area, demonstrating
  that all structures within this area are embedded in a single
  network of clusters, groups and filaments. The stellar mass density
  in the core of the SSC is always higher than $9{\times}10^{9}{\rm
  M}_{\odot}{\rm Mpc}^{-3}$, which is ${\sim}40{\times}$ the cosmic
  stellar mass density for galaxies in the local Universe. We find a
  new filamentary structure ($\sim$7\,Mpc long in projection)
  connecting the SSC core to the cluster A\,3559, as well as
  previously unidentified density peaks. We perform a weak-lensing
  analysis of the central 1\,deg$^2$ field of the survey obtaining for
  the central cluster A\,3558 a mass of
  $M_{500}=7.63_{-3.40}^{+3.88}\times10^{14}M_\odot$, in agreement
  with X-ray based estimates.
\end{abstract}


\section{Introduction}
\label{intro}

It is well established that the properties of galaxies are correlated
with the environment
\citep[e.g.][]{L02,GYF03,BEM04,BEH05,BBB06,PSE06,Haines07,BBW13}. At
$z{\simeq}0$ galaxy populations in rich clusters are dominated by
ellipticals, S0s and at lower masses dwarf ellipticals (dEs) with few,
if any, star-forming spirals \citep{D80,DTS85}. Cluster galaxies have
not always been as inactive as they are at the present
epoch. \citet{BO84} showed that the fraction of blue (star-forming)
galaxies among cluster members increases from almost zero in the local
Universe to $\sim$20\% by $z{\sim}0.4$, while recent Spitzer/Herschel
surveys have confirmed large numbers of starburst galaxies in clusters
to $z{\sim}1$ and beyond \citep[e.g.][]{PBR12}. While some of this
rapid evolution of cluster galaxies can be explained by the cosmic
${\sim}10{\times}$ decline in star formation among field galaxies
since $z{\sim}1$ \citep[e.g.][]{LPD05}, cluster populations have shown
an accelerated evolution in star formation (SF) over the last four
billion years, resulting in an overall ${\sim}15{\times}$ reduction in
the total star formation rates (SFRs) per unit halo mass since
$z{\sim}0.3$ \citep{HPS13}. Empirically, clusters accrete gas-rich,
star-forming spirals at $z{\ga}0$.5--1.0 and then contribute to
transforming them somehow into the passive S0s and dEs of local
clusters.

Several mechanisms affecting the galaxy properties and dependent upon
the environment have been proposed and investigated in detail and all
of them serve to kinematically disturb spiral galaxies and/or
transform their structural properties and/or deplete their reservoirs
of gas, and so quench star formation. These physical processes include
gravitational and tidal interactions amongst galaxies
\citep{TT72,MKL96}, between galaxies and the cluster gravitational
field \citep{BV90}, galaxy mergers \citep{BH91}, group-cluster
collisions \citep{B01}, ram-pressure \citep{GG72} and viscous
stripping \citep{N82}, evaporation \citep{CS77} and `starvation'
\citep{LTC80}. Since these mechanisms are characterized by different
time-scales and efficiencies which depend, in turn, on the properties
of both the galaxies themselves (e.g. their stellar masses,
morphologies) and their environment, they can affect the galaxy
properties in different characteristic ways
\citep{BG06,Haines07}. Examples of this are the different effects of
tidal and hydrodynamical interactions. The ram pressure exerted by the
hot and dense intracluster medium (ICM) can effectively remove the
cold gas supply, truncating the gas disc and quenching star formation,
in massive cluster galaxies passing through the centre of rich
clusters in about one crossing time ($\sim 10^9$\,yr), while the
interstellar gas of a massive galaxy may never be completely stripped
if the galaxy moves on a tangential orbit or is member of a poor
cluster. Repeated high-velocity encounters (harassment) of cluster
galaxies can destroy the fragile disc of dwarf galaxies, but not
significantly affect the structure of a giant galaxy.

The development of large spectroscopic surveys such as the Sloan
Digital Sky Survey (SDSS), plus the availability of panoramic far
ultra-violet -- far infra-red (FUV--FIR) data from the {\em GALEX} and
{\em Spitzer} space telescopes have allowed the impact of environment
on SF to be quantified in unprecedented detail. In these studies,
however, the environment and its characteristics are usually treated
as `static parameters'. While the galaxies move across the cluster,
field and filament environments (experiencing different ram pressures,
encountering other galaxies and possibly being involved in merging),
the cluster potential well, the galaxy density as well as the ICM are
usually not considered `time-dependent'. Of course, at first pass this
approach is an unavoidable reduction on complexity. On the other hand
in a hierarchical Universe with the assembly of the structures, the
galaxies evolve and move, tending towards denser regions with time,
while the environments change too, thus what we actually observe is
{\it galaxy evolution in an evolving environment}.

%GALAXY PROPERTIES AND ENVIRONMENT (VERY SHORT AND VERY SHARP)

The most massive structures in the local Universe are superclusters,
which are still collapsing with galaxy clusters and groups frequently
interacting and merging, and where a significant number of galaxies
are encountering dense environments for the first time. The relative
dynamical immaturity of superclusters and the presence of infalling
dark--matter halos make them ideal laboratories to test the
predictions of hierarchical mass assembly models, and in particular on
galaxy evolution. Superclusters are not so rare systems in the
Universe \citep[see][]{SD11}. The observations of superclusters is
often considered a challenge to the hierarchical structure formation
paradigm since such extreme dense structures, but also voids, are not
reproduced by the $N$-body simulations. \citet{YBA11} pointed out that
the reason of this discrepancy can be due to the method used to assess
the probability of finding such events in the distribution of cold
matter. They proposed a new technique to analyse an ensemble of
$N$-body simulations in a volume equal to that of the two-degree Field
Galaxy Redshift Survey \citep[2dFGRS,][]{C01} where the probability to
find peculiar structures (overdense and underdense) was thus estimated
to be $\sim$2 per cent and not null.

%WHY THE SUPERCLUSTERS TO STUDY GALAXY EVOLUTION?
%HOW MUCH ARE RARE THE SUPERCLUSTERS?

Superclusters allow in principle both to study dynamical processes
such as cluster-cluster collisions and group-cluster mergers and to
sample different environments from cluster cores to filaments and
fields. Furthermore, within a dynamically active and locally dense
structure the probability to observe evidence of environmental effects
on galaxy evolution is dramatically enhanced making these systems a
sort of {\it magnifying glass} to identify the different physical
mechanisms which transform the properties of galaxies. In order to
address all these aspects, a careful selection of the target is
fundamental. The ideal structure should map different environments
with evidence of cluster-cluster interactions. Finally, in order to
study in detail the galaxy properties a resolution $\lesssim 1$\,kpc
is required in a wide range of galaxy mass down to the dwarf regime
where such galaxies are not quenched by internal processes, but are
more susceptible to environmental transformations. With all this in
mind, we have undertaken a study of the Shapley supercluster which is
the largest conglomeration of Abell clusters in the local Universe.

The Shapley Supercluster Survey (ShaSS) will map a 23\,deg$^2$ region
($\sim$ 260\,Mpc$^{2}$) of the Shapley supercluster at $z{=}0.048$,
containing filaments and embedded galaxy groups which form a
dynamically-bound network connecting nine Abell and two poor clusters,
in order to identify the primary locations (groups, filaments,
clusters) and mechanisms for the transformation of spirals into the
S0s and dEs. 

Among the observational studies of superclusters, the
STAGES project \citep{G09} addressed in particular the study of galaxy
evolution in the Abell\,901(a,b) supercluster at redshift $z\sim$0.165
while the ORELSE survey (Lubin et al. \citeyear{LGL09}, but see also
Mei et al. \citeyear{MSH12}) searched for structures on scales greater
than 10\,Mpc at higher redshifs (0.6$<z<$1.3) with the aim to
investigate the properties of member galaxies. The comparison between
the STAGES, ORELSE and other similar studies and ShaSS will be then
unavoidable and useful to trace the evolution with redshift, although
we notice that the area (in Mpc$^2$) of ShaSS is a factor 10 that of
the STAGES survey and was chosen to map the filaments connecting the
Abell clusters.

%CRUCIAL IS THE TARGET CHOICE

The optical survey VST-ACCESS in four bands collected at the ESO VLT
Survey Telescope (VST) represents the core of this multi-band
project. The infrared coverage with WISE, the dedicated spectroscopic
survey with AAT/AAOmega and the Shapley-VISTA survey, together with
other proprietary data, provide the fundamental data-set to achieve
the scientific goals of ShaSS. In this article we will give an
overview of the project and present the first results. In
Sect.~\ref{objec} the motivations and main scientific objectives of
the project are discussed. The target is described in
Sect.~\ref{SSC}. The characteristics and strategy of the survey are
described in Sect.~\ref{survey}. Details of the data reduction and
analysis of the data quality are given in Sects.~\ref{DR} and
\ref{DA}, respectively. The first results concerning the
characterization of the environment are presented in
Sect.~\ref{results}, where we derive the galaxy density across the
whole ShaSS region and the underlying dark matter distribution for the
central 1\,deg$^2$ field which allows to estimate the mass of the
galaxy cluster A\,3558. In Sect.~\ref{sum} we present the summary and
conclusions.

Throughout the paper we adopt a cosmology with $\Omega_M$=0.3,
$\Omega_\Lambda$= 0.7, and H$_0$=70\,km\,s$^{-1}$Mpc$^{-1}$. According
to this cosmology 1\,arcsec corresponds to 0.941\,kpc at $z$=0.048 and
the distance modulus is 36.66. The magnitudes are given in the AB
photometric system.

%PAPER DESCRIPTION

\section{Survey motivations and objectives}
\label{objec}

\subsection{The role of large scale mass assembly on galaxy evolution}

In $\Lambda$CDM cosmological models structure formation occurs
  hierarchically, such that the most massive halos corresponding to
  galaxy clusters form latest, doubling their masses on average since
  $z{\sim}0.5$ \citep{BSW09,GNF12}, and are also the most dynamically
  immature. Preferentially located at the nodes of the complex
  filamentary web, clusters are continually accreting dark matter
  halos containing individual galaxies or galaxy groups. On average,
  50\% of galaxies in local clusters have been accreted since
  $z{\sim}0.4$ \citep{BSB09}, of which 40\% are within groups
  \citep{MBB09}.

The process of central galaxies becoming satellites, as their host
halos are accreted into more massive halos, has been shown to strongly
affect their evolution, with satellite galaxies more likely to be
quenched than central galaxies of the same stellar mass
\citep[e.g.][]{WTC12}.  Star formation is suppressed within galaxy
groups, with the fraction of star-forming galaxies declining steadily
with increasing group mass (at fixed stellar mass) and proximity to
the group centre \citep{WVY06,WTC12,WDF13}. Galaxies which are
quenched within such groups and later accreted into clusters are
described as having been `pre-processed'.

Numerical simulations and theoretical studies predict the effect of
cluster-cluster mergers on galaxy properties. \citet{BOC10} claimed
that merging of galaxy clusters may induce SF due to the
increase of the external pressure of the ICM compressing the cold gas
in the cluster galaxies. This results in a population of starburts or
post-starburst galaxies having the same age and so dating the merging
event. The spatial distribution of these star-forming galaxies is
expected to differ from the overall distribution of the other cluster
galaxies. Similarly, for comparable group and cluster halo densities
gravitational shocking as the group enters the cluster could
temporarily increases its mass and pull the group members into a
denser, more compact system inducing galaxy mergers \citep{Moss06}.
Simulations also showed that starbursts induced by galaxy
mergers can be amplified by a factor $\sim 2$ \citep{MB08} if they
take place in the tidal field of cluster/group of galaxies.

\citet{OLK05} studied the excess population of radio galaxies in the
cluster A\,2125 at $z$=0.25, mostly located in groups outside the
cluster core and with radio luminosities indicating the SF as the main
mechanism responsible of the radio emission. The authors related this
observational evidence with the ongoing major cluster-cluster merger
and explained the increased SF as due to the variation of the tidal
field experienced by the member galaxies during a cluster-cluster
interaction, probably being close to the core passage. Studying a
system of merging clusters, \citet[][]{JHS08} observed an excess of
star-forming galaxies aligned along the bridge of galaxies connecting
A\,3158 to the A\,3125/A\,3128 complex, suggestive of merger-induced
SF. In addition, they found that the fraction of radio-loud sources is
lower with respect to a global cluster environment and similar to that
measured by \citet{VBM00} in the cluster A\,3558 which is considered
as an example of a late merger stage. They suggested that in the
cluster cores radio emission from active galactic nuclei (AGNs) is
suppressed in the late stages of cluster mergers. The suppressed radio
emission observed in both bright and faint cluster galaxies has been
also associated to cluster merger by \citet{MM07}. They explained
their findings with the disruption of cool cores, i.e. the gas supply
to the central AGN for the bright galaxies, while the low radio
loudness and fainter cluster galaxies have been affected by the
enhanced ram pressure when crossing the shock front between the
merging clusters. Further observational support of the important role
that cluster mergers play in triggering the evolution of cluster
galaxies is provided by \citet{OCN12} studying the merging cluster
A\,2744. They identified three rare `jellyfish' galaxies located in
close proximity to the ICM features associated with a merging
subcluster and its shock front. Their interpretation is that the SF
knots detected in the tails of stripped gas are due to the rapid
increase in pressure experienced by the galaxies interacting with the
shock.

Hence, the observations seem to confirm that galaxy/group accretion
and cluster-cluster merger affect the cluster galaxies and that the
galaxy properties can be sensitive indicators of the merger stage.

\subsection{Measuring the environmental effects beyond r$_{200}$} 
\label{r200}

The colour-density relation is observed in galaxy clusters well beyond
$r_{200}$ \citep[e.g.][]{Haines09} and the fraction of star-forming
galaxies steadily increases with cluster-centric radius. Nevertheless
even at 3-5$\times$ r$_{200}$, the fraction of star-forming galaxies
is below that seen in the field \citep[e.g.][]{CEG11} suggesting that
the environment plays a role inducing transformations well outside
$r_{200}$.

By means of hydrodynamical cosmological simulations \citet{BMB13}
investigated the trend of cold and hot gas contents and SF as function
of clustercentric distance in clusters and groups of galaxies. They
found a large-scale trend with cold and hot gas contents and SF
increasing with the clustercentric distance, but approaching the
values of the field galaxy sample only at $\sim 5r_{200}$
(corresponding to 10\,Mpc for a massive cluster). Moreover, the SF in
low-mass cluster galaxies (M=10$^{9-9.5}$M$_{\odot}$) still shows a
significant discrepancy even at this large radius with respect to that
of the field galaxies. According to the authors, three main causes can
explain the observed trends: i) `pre-processing' of galaxies within
infalling groups; ii) `overshooting' for those galaxies that are not
falling in for the first time (also called `backsplash' galaxies);
iii) ram-pressure stripping. Of course, the three mechanisms act
differently depending on the galaxy and host halo mass and the
clustercentric distance, but in any case the radial trend of the hot
gas cannot be explained solely by overshooting and pre-processing. A
direct interaction with the host group/cluster is also required and
their simulations suggest that ram-pressure stripping can strip
the hot gas from low- and high-mass galaxies out to several times
$r_{200}$, although it cannot affect the cold gas except for the
low-mass galaxies. The authors also claimed that tidal interaction
does not play an important role in the gas removal, but it should be
an important factor in the morphological transformation of these
galaxies.

S0s differ from normal spirals by their higher bulge luminosities
rather than fainter disks \citep{CZ04}, disfavouring simple RPS or
starvation mechanisms \citep[but see][]{KSS09}. Instead, mechanisms
such as merging or harassment are capable of channelling material to a
central bulge, sufficient to produce the higher central mass densities
seen in cluster spirals and ultimately the stellar phase densities
found in S0s \citep{MMT07}. This channeling of material will also
likely fuel rapid growth of the central supermassive black hole and
trigger a period of nuclear activity. Deep observations are needed to
reveal tidal streams or `fans', characteristic of recent merging
activity \citep{vD05}.

In-situ transformation via merging in clusters is strongly suppressed
due to the high encounter velocities. Low-velocity encounters and
mergers should be much more frequent in galaxy groups, leading to the
suggestion that the bulge growth required to form S0s occurs primarily
via pre-processing in galaxy groups, which are subsequently accreted
into clusters \citep{BEM04}. Moreover, the S0 fractions of $z{\la}0.5$
groups match those seen in clusters at the same redshifts, and are
much higher than found in the field \citep{WOM09}. Strong nuclear
activity which is linked to bulge growth is also found to be
effectively suppressed in cluster galaxies, with the distribution of
X-ray AGN in rich clusters revealing them to be an infalling
population with high relative velocities \citep{HPS12}. This confirms
that the transformation of galaxies often occurs (or begins at least)
well outside the cluster cores, in groups or as infalling galaxies
enter the virialized region for the first time \citep{MET05}.

Hints about the distribution of galaxy luminosity, colours and
morphology in the filaments have been given by \citet{PB06}, who found
that brighter galaxies have a less filamentary distribution than the
fainter ones. With respect to the morphology, the early-type galaxies
are concentrated in the vicinity of the nodes, while spiral galaxies
are sparsely distributed across the filaments. \citet{FBM08} observed
that the fraction of starburst galaxies in the filaments around the
cluster A\,1763 is twice that in other cluster regions as detected by
{\it Spitzer} and suggested that filaments are a sort of {\it galaxy
  reservoir} for clusters.

By studying a sample of supercluster filaments in the two-degree Field
Galaxy Redshift Survey, \citet{PRP08} observed a sudden enhancement in
SF in faint dwarf galaxies outside the cluster virial
radius. They interpreted their findings with a close interaction
and/or harassment with other infalling galaxies along the same
filaments. Another possible explanation is the very first interaction
that these galaxies experienced with the ICM inducing tidal shocks and
then a burst of SF. Actually, the dominant processes that
quench SF in galaxies depend crucially on the galaxy
mass. Dwarf galaxies, given their shallow potential wells, should be
more susceptible to environmental processes such as tidal/ram-pressure
stripping \citep[e.g.][]{BWG09}. Indeed their SF histories are
completely defined by their local environment, since passive dEs are
only found as satellites within massive halos \citep[i.e. cluster,
group or massive galaxy;][]{Haines06,Haines07}. The formation of many
cluster dEs is often rather recent, manifest by young stellar ages
(${\sim}2$\,Gyr) and the significant populations of starburst and
post--starburst dwarf galaxies in the outskirts of local clusters
\citep{SLH09,MHR11}.

\subsection{Relating star formation quenching to ram-pressure stripping}
\label{RPS}

Ram-pressure stripping (RPS), as originally proposed by \citet{GG72}
requires, in principle, the presence of a dense ICM. Thus, its effect
would be limited to cluster cores where the gas discs of massive
spirals are rapidly truncated. Nevertheless, with a 3--D
hydrodynamical simulation, \citet{MBD03} showed that RPS may extend to
poorer environments for low-mass galaxies which, thanks to their lower
escape velocities, are easier to strip. \citet{RH05}, by means of high
resolution 2D hydrodynamical simulations, demonstrated that ram
pressure effects can be observed over a wide range of ICM
conditions. In particular, in high density environments RPS severely
truncates the gas disc of L$^\star$ galaxies, while in low density
environments, where moderate ram pressure is foreseen, their gas disc
is clearly disturbed and bent \citep{RH05}. The gas discs of these
galaxies can be truncated to 15-20\,kpc in the first 20--200\,Myr of
RPS.

\begin{figure*}
\includegraphics[width=150mm]{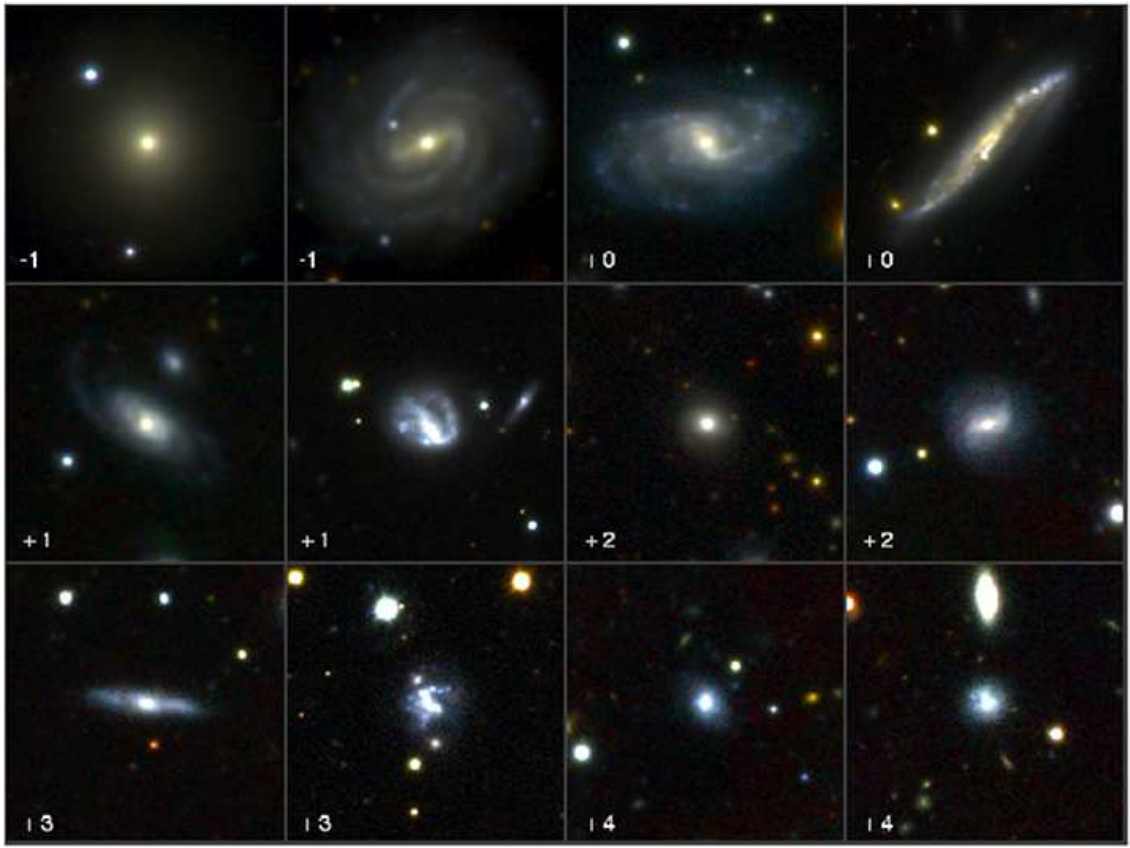} 
\caption{VST-ACCESS {\it gri} composite images of a sample of galaxies
  spectroscopically confirmed as members of the Shapley
  supercluster. From left to right and top to bottom different
  magnitude bins from m$^\star$-1 to m$^\star$+4 as indicated. Each
  square frame is 1\,arcmin ($\sim 56$\,kpc) wide.}
\label{gals}
\end{figure*}

\citet{ACCESSV} identified a bright (L$>$L$^\star$) barred spiral
galaxy 1\,Mpc from the centre of the rich cluster A\,3558 in the
Shapley supercluster core, which is strongly affected by
RPS. Integral-field spectroscopic observations revealed ongoing gas
stripping in the form of one-sided extraplanar ionized gas along the
full extent of the disk, simultaneously with a starburst triggered by
gas compression along the leading edge of the galaxy. The galaxy is
estimated to be being subjected to weak-moderate ram pressure, as
defined by \citet{RH05}. This adds a piece of evidence to the fact
that RPS is acting more efficiently on the galaxy interstellar medium
(ISM) than previously foreseen and also outside of the cluster cores
as also observed in the Virgo cluster by \citet{CGK09}. This new
understanding of the RPS supports the view that this mechanism is the
principal transformation process to quench SF in spirals although
probably helped by other processes affecting the structure of the
galaxies. \citet{BBW13} drew a similar conclusion analysing a sample
of 182 disk galaxies in the cluster system Abell\,901/902. The
fraction of galaxies showing asymmetric gas rotation curves, and thus
probably affected by RPS \citep{KKU08}, turned out to be 75 per cent
higher in the clusters than in the field, the majority of them being
morphologically undisturbed. Although these galaxies seemed
preferentially located at low clustercentric radii, they also observed
a population of dusty and red spirals in the cluster outskirts, which
could be also affected by ISM-ICM interactions. The relevance of RPS
for quenching SF in cluster galaxies is further complicated by the
fact that a non-homogeneous ICM is expected for non virialized merging
and post-merging clusters, thus the clustercentric distance is not the
only parameter describing the ICM density, i.e. the ram-pressure
strength.

\subsection{The approach of the Shapley Supercluster Survey}
\label{HOW}

Our study is based on characterizing a dynamically active environment
in a multifaceted way and disentangling its effects on galaxy
evolution. It has the following main objectives.

\begin{description}
\item [-] To investigate the role of cluster-scale mass assembly on
  the evolution of galaxies mapping the effects of the environment in
  the cluster outskirts and along the filaments with the aim to
  identify the very first interactions between galaxies and their
  environment.
\item [-] To identify and measure signs of ongoing transformation in
  galaxies belonging to a complex structure with the goal of improving
  our comprehension of what drives their star-formation quenching and
  structural modification.
\item [-] To obtain detailed maps of the dark matter and baryonic
  matter distributions (galaxies, ICM), combining weak lensing, X-ray
  and dynamical analyses.
\item [-] To quantify the variation in the stellar mass fractions
  going from cluster cores to groups, by comparing the near-infrared
  light distribution with the dark matter maps and dynamical masses.
\item [-] To build up a multi-band homogeneous data-set on this area
  of the sky made of sub-kiloparsec resolution imaging and
  magnitude-limited spectroscopy, thus providing the community with a
  solid background for studies of the Shapley supercluster.
\end{description}

To address the above objectives we will explore the global and
internal properties of galaxy populations extending outside the
cluster/group virial radius and aim for an accurate characterization
of the environment. This will be defined through galaxy density, dark
matter distribution, dynamical substructure, and ICM properties. The
different quantifications of the environment will allow us to
disentangle the effects of local and large-scale density, cluster and
group merging, dynamical state and mass of the host system on the
properties of galaxies in different ranges of mass.

Recent studies investigated the effects of the local and large-scale
environment on galaxy evolution, mainly considering groups of
galaxies. Some of these works defined the environment through the
luminosity-density only \citep[e.g.][]{LTH12}, or associate galaxy
groups to a large-scale environment identified by a smoothed
luminosity field \citep[e.g.][]{LLY13}, or are limited to massive
galaxies \citep[e.g.][]{CSC13}. Therefore, although there is a general
agreement that the group environment is important for the evolution of
the group members, a study of global and internal properties of galaxy
population in groups and, in general, in different structures
dynamically bound in a supercluster is still lacking. For instance,
\citet{RMB12} found that both local (distance from the centre) and
global (mass) group environments play a role in quenching star
formation, while \citet{Ziparo14} show that global group-scale
mechanisms linked to the presence of a hot gas halo are dominant in
quenching SF in group galaxies rather than purely density related
processes. 

In the following we briefly outline how we will derive the quantities
necessary for our study (see also Sect.~\ref{survey}). 

As the main classification scheme of galaxies we will adopt the
concentration, asymmetry and clumpiness parameters \citep{K85,
  BJC00,CBJ00,C03}, complemented by the M20 and Gini coefficients
\citep{LPM04,AvN03}. This parameter set (hereafter `CAS+MG') has
proven to robustly link the internal light distribution of galaxies to
their formation and evolution \citep[e.g.][and references
therein]{SCL07,MGZ09,KPO14,HMC14,LJC08,LJC11}. Besides being effective
in separating the different morphological types and tracing star
formation, the CAS+MG scheme is particularly sensitive to their recent
interaction or merging activities, making it the ideal tool to obtain
a census of galaxies whose structure appears disturbed by the
environment, which is crucial for our project. For a selected sample
of `normal' galaxies (as determined through the CAS+MG parameters) it
will be possible to quantify the relevance of (pseudo-)bulges and bars
as a function of the environment by surface photometry fitting.

In Fig.~\ref{gals} we show examples of SSC galaxies in the magnitude
range from m*-1 to m*+4 ($r=14-19$\,mag). For galaxies in this
magnitude range we plan to perform classical (visual) morphological
classification, that will be used to calibrate the CAS+MG parameters
against the Hubble sequence for our data. The signal-to-noise ratio
(SNR) and spatial resolution of VST images will instead allow us to
obtain reliable CAS+MG parameters up to m*+6 ($r=21$\,mag), i.e. well
into the dwarf galaxy regime (see Sect.~\ref{optical}). Finally,
AAOmega and literature spectra of more luminous galaxies will provide
spectral classification through line indices.

Galaxy global properties such as colours, stellar masses and
star-formation rates will be derived. The contribution of both
unobscured and obscured SFR need to be considered since dust-obscured
cluster galaxies are common in merging clusters \citep[e.g.][]{HSE09}.
The $u$--band luminosity will provide a star-formation rate (SFR)
indicator once calibrated using the multi--wavelength data already
available in the Shapley supercluster core (see Sect.~\ref{CD}), by
obtaining correlations between luminosity, dust extinction, and
metallicity. The available mid-infrared data will provide a robust and
independent indicator of obscured star formation (Sect.~\ref{CD}), and
also allow us to identify AGN via their unusually red W1-W2 colours
\citep[e.g.][]{WISE10}. AGN will also be identified from their
emission-line ratios or broad emission lines from the AAOmega spectra,
or their X-ray emission in the supercluster core where XMM data are
available.

The depths of our survey are conceived to reach the necessary accuracy
of all of the above quantities, as will be outlined in the next
Sections.

Finally, we will not only measure statistically the environmental
effects on galaxy properties on such large scales, but also to `catch
in the act' the direct interaction of supercluster galaxies with their
environment into the surrounding large-scale structure. The latter is
possible only using imaging with a sub-kiloparsec resolution and
follow-up integral-field spectroscopy for a few individual cases
\citep[e.g.][]{ACCESSV}.

\section{The target: Shapley supercluster}
\label{SSC}
The `remote cloud of galaxies in Centaurus' first identified by
\citet{S30} is one of the richest superclusters in the nearby
Universe, consisting of as many as 25 Abell clusters \citep{ZZS93} in
the redshift range 0.033$< z <$0.060 \citep{QRM95,QMP97}. The first
spectroscopic study confirming the existence of the `Centaurus
supercluster' was carried out by \citet{MM87}, then it was
re-discovered by \citet[][their `$\alpha$-region']{SBC89} as a cluster
overdensity in the \citet[][ACO]{ACO89} cluster catalogue and
identified by \citet[][his `Shapley concentration']{R89} as an excess
in galaxy number counts in the UK Schmidt Telescope Sky Survey plates.

The Shapley supercluster (hereafter SSC) because of its peculiar
richness and location, lying in the direction of the dipole anisotropy
of the Cosmic Microwave Background (CMB), was investigated as
responsible of at least a fraction of the Local Group acceleration
\citep[][and references therein]{SBC89,R89,PV91,QRM95,KME04,FKK13}. To
assess the role of this structure in the acceleration of the Local
Group a robust estimate of its mass is required. This gave a boost to
spectroscopic and X-ray observations of the supercluster devoted to i)
map the whole structure and measure the density contrast in galaxy
number counts and mass/number of associated clusters; ii) investigate
its dynamical state and iii) derive the underlying mass
distribution. A detailed review of the properties of the Shapley
supercluster based on previous investigations will be given elsewhere
(Merluzzi et al., in preparation), here we will describe the key
characteristics of the SSC - i.e. those features making the SSC, and
above all its central region, {\it a magnifying glass} to investigate
the effect of the environment and mass assembly on galaxy evolution.

\subsection{Supercluster morphology and dynamics}
By analysing a spectroscopic sample across a region of 15\,deg in
diameter centred on the dominant cluster A\,3558, \citet{QRM95}
concluded that the supercluster has a `cigar-shape with the eastern
side being the closest to us' \citep[but see also][]{DPP04} and from
the flattened geometry they suggested that it is not spherical and
virialized. The complex morphology of the SSC comprises a main body at
$cz\sim$15000\,km\,s$^{-1}$ together with walls/filaments/arms of
galaxies connecting the three main systems of interacting clusters
(the A\,3558, A\,3528 and A\,3571 complexes) as well as a foreground
structure connecting the SSC to the Hydra-Centaurus supercluster
($cz\sim$4000\,km\,s$^{-1}$) and hints of a background structure at
$cz\sim$23000\,km\,s$^{-1}$\citep[for details see][]{QCR00,PQC06}. The
main plane of the SSC ($cz\sim$14800\,km\,s$^{-1}$) extends
10x20\,deg$^{2}$ \citep[see][]{DPP04}. Is this vast and complex
structure gravitationally bound? In this case it would be the most
massive bound structure known in the Universe.

Applying a spherical collapse model, \citet{RQC00} found that the SSC
is gravitationally collapsing at least in its central region within a
radius of 8$h^{-1}$\,Mpc centred on A\,3558 including 11 ACO clusters;
the very inner region, associated with the massive clusters, is likely
in the final stages of collapse. X-ray observations of clusters in
the SSC confirmed the overdensity: a factor 10 to 50 cluster
overdensity \citep{RFE91} and factor 3 baryon overdensity \citep{F91}
over a region of 60x80\,Mpc$^{2}$. These results indicated that the
structure is gravitationally bound at least in its central region.

From the ROSAT All-Sky Survey in a region of 0.27\,sterad,
\citet{DSE05} measured the cluster number density which turned out to
be more than an order of magnitude greater than the mean density of
Abell clusters at similar latitudes - mainly due to an excess of
low-luminosity X-ray systems in the outskirts, suggesting that the
supercluster is still accreting low-mass systems. \citet{EFW97}
analysing a mosaic of 15x20\,deg$^{2}$ of ROSAT and Einstein
Observatory X-ray observations centred on A\,3558, concluded that the
SSC core is dynamically bound within $\sim 9$\,Mpc, approaching the
maximum expansion before collapsing, i.e. the turnaround point. On the
other hand, the dynamical analysis of \citet{BZZ00} proposed a
scenario where the SSC has already reached its turnaround radius and
the final collapse will happen in $\sim 1$\,Gyr.

Using the X-ray cluster sample of \citet{DSE05} and the spherical
collapse model, \citet{ML08} investigated the dynamics of the
SSC. Their study showed that the SSC is not bound at a radius of
51\,Mpc with an outer shell moving radially away from the centre,
while the excess of mass becomes enough to bind a spherical region of
$\sim$10\,Mpc radius.

Recently, \citet{PB13} ran $N$-body numerical simulations of the SSC
and other superclusters in order to determine what portions of the
superclusters were potentially gravitationally bound. The SSC showed
the most extended bound structure among the other analyzed
superclusters. In particular, A\,3554, A\,3556, A\,3558, A\,3560 and
A\,3562 have a large number of close encounters in their simulations,
while an additional pair, A\,1736 and A\,3559, has also some chance of
being bound. In such a crowded environment it is very unlikely that
galaxies have not been affected by cluster related processes such as
RPS, starvation and/or galaxy harassment and tidal
interaction.

One of the key objectives of ShaSS is to constrain the dark matter
distribution and mass over this whole region via a weak lensing
analysis, and in conjunction with X-ray and dynamical analyses
estimate the extent of the region that is gravitationally bound to the
SSC core, and determine whether clusters such as A\,3559 are currently
falling towards the supercluster core.

\subsection{The Shapley supercluster core}
\label{SSCC}
We now zoom in to the very central dense region of SSC, namely the
`core', which consists of three Abell clusters: A\,3558, A\,3562, and
A\,3556 and two poor clusters (SC\,1327-312 and SC\,1329-313),
resolved by \citet{BRF94} by means of X-ray observations. The
evolutionary stage of the SSC core (SSCC) is still matter of
debate.
%THE CORE 

\begin{table*}
  \centering
\caption{{\bf Galaxy clusters in the ShaSS region.}}
  \begin{tabular}{lllllcccllll}
    \hline \hline 
{\bf Cluster} & {\bf RA} J2000 & {\bf DEC} J2000 & \multicolumn{2}{l}{{\bf redshift}} & \multicolumn{2}{c}{{\bf $\sigma$} [km\,s$^{-1}]$} & {\bf richness}$^a$ & \multicolumn{2}{l}{{\bf R$_{VIR}$} [Mpc]} & \multicolumn{2}{l}{{\bf Mass}$^b$ [M$_\odot$]} \\ 
\hline
AS\,0724 & 13 13 08.6 & -32 59 38\,$c$ & 14864$\pm$157 & $d$ & 510$\pm$85 & $d$ & 0 & & & 8.8$\times$10$^{13}$ & $e$ \\

AS\,0726 & 13 15 11.7 & -33 38 52\,$d$ & 14892$\pm$137 & $d$ & 578$\pm$77 & $d$ & 0 & 0.96 & $f$ & 4.6$\times$10$^{13}$ & $f$ \\

A\,3552 & 13 19 00.7 & -31 51 04\,$c$ & 14753$\pm$119 & $d$ & 682$\pm$60 & $d$ & 1 & & & 3.6$\times$10$^{13}$ & $e$ \\

A\,3554 & 13 19 27.6 & -33 29 49\,$c$ & 14431$\pm$94 & $d$ & 560$\pm$66 & $d$ & 1 & 0.80 & $f$ & 5.8$\times$10$^{13}$ & $f$ \\

A\,3556 & 13 24 00.2 & -31 39 22\,$c$ & 14357$\pm$76 & $g$ & 643$^{+53}_{-43}$ & $g$ & 0 & 0.98 & $f$ & 1.7$\times$10$^{14}$ & $f$ \\ 

A\,3558 & 13 28 02.6 & -31 29 35\,$c$ & 14403$^{+60}_{-55}$ & $g$ & 996$^{+40}_{-36}$ & $g$ & 4 & 1.16 & $f$ & 1.3$\times$10$^{15}$ & $g$ \\

A\,3559 & 13 29 53.1 & -29 30 22\,$h$ & 14130$\pm$57 &  $d$ & 519$\pm$45 & $d$ & 3 & 0.31 & $f$ & 2.0$\times$10$^{13}$ & $f$ \\

A\,3560 & 13 32 22.0 & -33 05 24\,$i$ & 14551$\pm$106 & $d$ & 793$\pm$116 & $d$ & 3 & 1.33 & $f$ & 3.4$\times$10$^{14}$ & $f$ \\

A\,3562 & 13 33 47.0 & -31 40 37\,$c$ & 14455$\pm$191 & $d$ & 1197$\pm$194 & $d$ & 2 & 0.89 & $f$ & (3.9$\pm 0.4$)$\times$10$^{14}$ & $j$ \\

SC\,1327-312 & 13 29 45.4 & -31 36 12\,$k$ & 14844$^{+105}_{-211}$ & $g$ & 691$^{+158}_{-146}$ & $g$ && 1.30 & $f$ & 3.0$\times$10$^{13}$ & $f$ \\

SC\,1329-313 & 13 31 36.0 & -31 48 45\,$k$ & 14790$^{+114}_{-67}$ & $g$ & 377$^{+93}_{-82}$ & $g$ && 1.14 & $f$ & 3.7$\times$10$^{13}$ & $f$ \\
&&& 13348$^{+69}_{-83}$ & $g$ & 276$^{+70}_{-61}$ & $g$ &&&& \\
\hline
\hline
\end{tabular}
\begin{tabular}{l}
a) \citet{ACO89}. \\
b) All the masses are dynamically derived except for that of A\,3562 which is derived from X-ray observations. \\
c) \citet{DSE05}. \\
d) \citet{PQC06}, the redshift is corrected with respect to the CMB. \\
e) \citet{RQC00}, cluster mass within a radius enclosing an average density 500 times the critical density. \\
f) \citet{RMP06}, estimates of virial radius and virial mass. \\ 
g) \citet{BZZ98}, the mass value is transformed into adopted cosmology. \\
h) \citet{DFJ99}. \\ 
i) \citet{BVZ02}. \\
j) \citet{EBD00}, the mass value is transformed into adopted cosmology. \\ 
k) \citet{BRF94}. \\
For SC\,1329-313 mass and virial radius are estimated for the whole
system, while redshifts and velocity dispersions \\ are given for the
two different clumps by \citet{BZZ98}. \\
Uncertanities are quoted when available. \\
\end{tabular}
\label{clusters}
\end{table*}

The dynamical analysis of the SSCC pointed out its very complex
dynamical state with interacting clusters belonging to the same
structure elongated both in declination and along the line of sight
\citep{BZV94,BZZ98}. Several sub-condensations, detected also in X-ray,
could be recognized. \citet{BPR98} identified twenty-one significant
three-dimensional subclumps, including eight in the A\,3558 cluster
alone. This `clumpy' structure as well as the proximity of the
clusters (e.g. the Abell radius of A\,3556 is overlapped to that of
A\,3558) makes robust estimates of the cluster velocity dispersions,
masses \citep[see][]{BZZ98} and even Abell richness
\citep[see][]{MGP94} difficult. Based on their spectroscopic study,
\citet{BPR98} proposed two alternative evolutionary scenarios for the
SSCC: i) a cluster-cluster collision seen just after the first
core-core encounter; ii) a series of random mergings occurred among
groups and clusters.
%CORE DYNAMICS

Diffuse filamentary X-ray emission has been observed across the whole
SSCC \citep{BZM96,KB99,HTS99} connecting the clusters. A continuous
filament of hot gas connecting A\,3562 and A\,3558 was also seen in
the reconstructed thermal Sunyaev-Zeldovich (SZ) {\em Planck} survey
map (Planck Collaboration \citeyear{Planck13}). The distribution of
the X-ray emission was found to be clearly asymmetric in A\,3562,
A\,3558 and SC\,1329-313 and gradients of gas temperature and
metallicity have been measured \citep{EBD00,AKF03,FHB04} suggesting
that cluster-cluster mergers have occurred and/or are ongoing. By
analysing XMM-Newton observations, \citet{FHB04} proposed a tidal
interaction between SC\,1329-313 and A\,3562 to explain the observed
properties of the hot gas in A\,3562 -- the tailed shape of the X-ray
emission associated with SC\,1329-313 as well as the sloshing of the
A\,3562 core. Another detailed analysis of XMM-Newton and Chandra by
\citet{RGM05} pointed out the very complex dynamical history of
A\,3558 having characteristics which are typical of both merging
(e.g. gas temperature gradients) and relaxed (e.g. cool core)
clusters. They also detected a cold front leading in the NW direction
and probably due to the sloshing of the cluster core induced by the
perturbation of the gravitational potential associated with a past
merger.
%CORE X-RAY

  A weak and asymmetric radio halo has been detected in A\,3562
  \citep{VBD03,GVB05}. The radio halo correlates with the X-ray
  emission and presents a radio head-tailed galaxy embedded in it and
  located between the cluster core and SC\,1329-313, thus supporting
  the interaction between the two systems as proposed by
  \citet{FHB04}.
%CORE RADIO

  A deficit of radio galaxies with respect to the radio-optical
  luminosity function of other clusters, although probably to be
  ascribed to A\,3558 alone, was observed by \citet{VBM00}. On the
  other hand, \citet{M05}, with a radio survey of a 7\,deg$^{2}$
  region of the SSCC, found a dramatic increase in the probability for
  galaxies in the vicinity of A\,3562 and SC\,1329-313 to be
  associated with radio emission suggesting young starbursts related
  to the recent merger of SC\,1329--313 with A\,3562. This observation
  could be explained by galaxy merging efficiently transferring gas
  into the galaxy centre, feeding AGN and then switching on SF
  \citep{B99} .

  Although the studies mentioned above are fundamental to demonstrate
  the complex dynamical status of the SSC and its core, none of them
  could systematically tackle the issue of galaxy evolution in the
  supercluster environment due to the lack of accurate and homogeneous
  multi-band imaging covering such an extended structure. This
  prevented to collect information about the integrated (magnitudes,
  colours, SFR) and internal properties (morphological features,
  internal colour gradients) of the supercluster galaxies. Excluding
  observations of single clusters in SSC, the first CCD observations
  covering a $\sim 2$\,deg$^2$ contiguous area of SSCC were those of
  the ACCESS project \citep[PI: P. Merluzzi;][see
  Sect.~\ref{ACCESS}]{SOSI,SOSII} which analyzed a galaxy sample
  complete down to $B$=22.5 ($>m^\star$+6) and $R$=22.0
  ($>m^\star$+7), i.e. well into the dwarf galaxy regime (see
  Table~\ref{mstar}).

\subsection{ACCESS project: main results and open issues} 
\label{ACCESS}

The multi-wavelength data-set of the ACCESS \footnote{European funded
  project: {\it ACCESS: A Compete CEnsus of Star formation in the
    Shapley supercluster}, PI: P. Merluzzi; www.oacn.inaf.it/ACCESS}
project \citep[see][]{ACCESSI}, covering 2-3\,deg$^{2}$ of the SSCC
allowed us to obtain the complete census of stellar content and SF
across the core region from A\,3556 to A\,3562. This project was
  dedicated to investigating the effect of the environment on galaxy
  evolution in the SSCC exploiting one of the first multi-wavelength
  data-sets available for such a wide area of a supercluster. The data
  include panoramic imaging in the UV (Galaxy Evolution Explorer, GALEX),
  optical (ESO Wide Field Imager, WFI), NIR (UKIRT/WFCAM) and
  mid-infrared (Spitzer/MIPS), as well as high signal-to-noise ratio
  (S/N) medium-resolution optical spectroscopy (AAT/AAOmega) for 448
  supercluster galaxies. In the following we mention the main results
  of this project and the open issues that stimulated us to undertake
  a new multi-wavelength study over a wider supercluster area.

By studying the optical \citep{SOSI} and near-infrared
\citep[NIR,][]{ACCESSI} luminosity functions (LFs) down to
m$_K^\star$+6 it was found that the slope of the LF increases from
high- to low-density environments, indicating that mechanisms such as
galaxy harassment and/or tidal stripping contribute to shape the
LF. The stellar mass function (SMF), however, does not seems to change
its trend with galaxy density in the SSCC and does not show the sharp
upturn below $\mathcal{M}$=10$^{9.5}$M$_{\odot}$ observed in the field
galaxy population \citep{PBZ10,BDL12}. Is this difference due to the
different evolution of the supercluster and the field galaxies, or it
is only an artifact due to the different range of stellar masses
analyzed for field and supercluster galaxies?  And also, which are the
different contributions to the galaxy SMF of the blue and red galaxy
populations in the supercluster from the cluster cores to the
filaments and field? To answer these questions, the mass range should
be extended and the membership ascertained.

We also found evidence that the bulk of the star-forming galaxies have
been recently accreted from the field and have yet to have their
star-formation activity significantly affected by the cluster
environment \citep{ACCESSII} and that the vast majority of SF seen in
the SSCC comes from normal infalling spirals
\citep{ACCESSIII}. Nevertheless, this analysis was limited to galaxies
belonging to the SSCC and thus in an extremely dense environment. With
the aim to reach a comprehensive picture of how and where galaxies
start to quench their SF, it is important to move out from the cluster
cores and analyze the properties of the infalling galaxies, i.e. to
follow the cluster member from the {\it converging} filaments into the
clusters. This will also enable us to investigate wether the formation of
early-type galaxies, dominant in cluster cores, can be driven by {\it
  morphological quenching} \citep[see][]{MBT09} or RPS/starvation
\citep[e.g.][]{CK08}. It is unclear which is the most likely and
effective process at work \citep{ACCESSIV}. Wether the formation of
passive early-type galaxies in cluster cores should involve the {\it
  prior} morphological transformation of late-type spirals into S0/Sa,
mechanisms as pre-processing and tidal interaction have to be
considered and quantified. In order to understand if this is the case,
we need a morphological study of supercluster galaxies, both isolated
and in groups, in the cluster outskirts and beyond.

\begin{figure*}
\includegraphics[width=168mm]{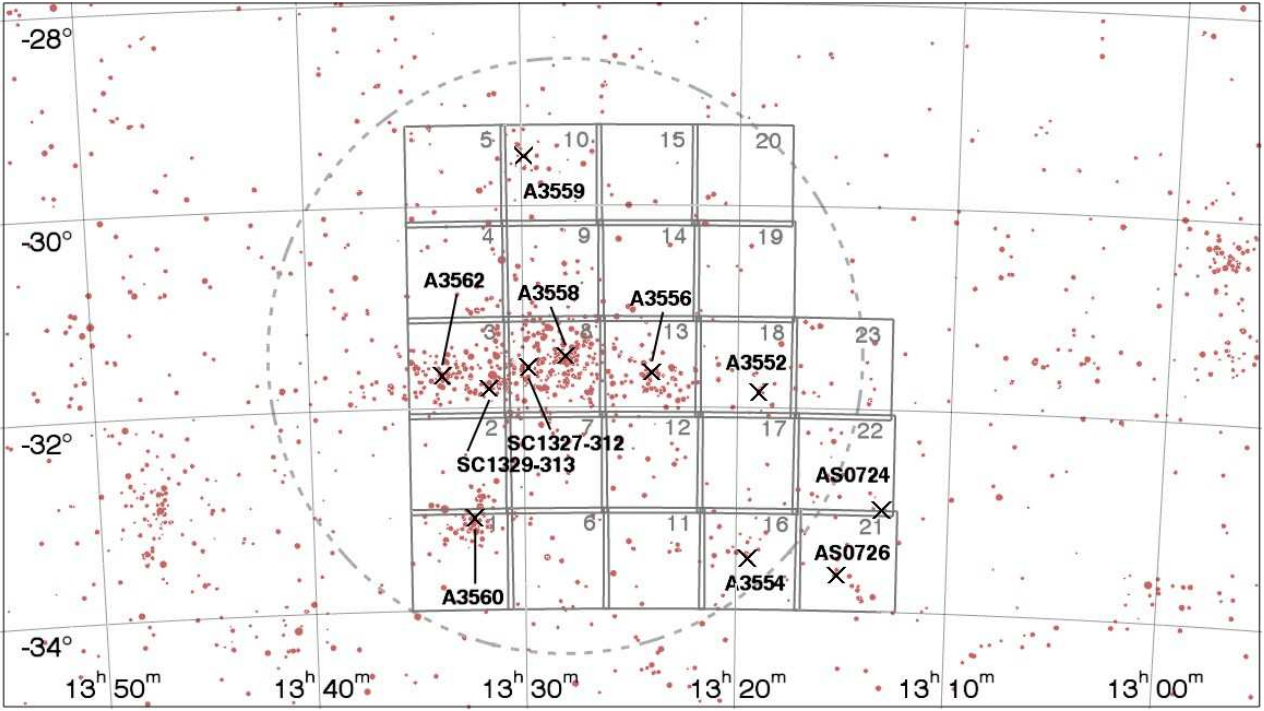} 
\caption{The 23 1\,deg$^2$ VST fields mapping the ShaSS region. Red
  dots indicate the supercluster members in the range
  $V_h$=11300--17000\,km\,s$^{-1}$ taken from literature. The size of
  the dots are proportional to the $K$-band flux. Black crosses show
  the cluster centres. The 10\,Mpc radius dotted circle encloses the
  supercluster region believed to be dynamically bound. The SSCC
  corresponds to fields \#3,8,13. The positions of all structures
  present in the plotted area and the given redshift range are indicated.}
\label{VST-ACCESS}
\end{figure*}

\section{ShaSS: the data}
\label{survey}

We will map a region of $\sim$260\,Mpc$^2$ including the SSCC and
other six galaxy clusters (AS\,0724, AS\,0726, A\,3552, A\,3554,
A\,3559 and A\,3560, see Fig.~\ref{VST-ACCESS}). The supercluster
region is chosen to ensure to map the structures directly connected to
the SSCC. In fact, the 11 clusters in the region are all within
500\,km\,s$^{-1}$ of the central cluster A\,3558. The survey
boundaries are chosen to cover all 11 clusters and the likely
connecting filaments, but also to extend into the field (fields 19 and
20 in Fig.~\ref{VST-ACCESS}). The main characteristics of the clusters
and groups in the survey are listed in Table~\ref{clusters}. We would
like to point out that among the other quantities taken from the
literature, the virial radii and masses should be considered only
indicative for the SSC clusters studied here for which the assumptions
of spherical symmetry and isotropy of the velocities are likely not
applicable.

The data-set includes optical ($ugri$) and NIR ($K$) imaging
  acquired with VST and VISTA respectively, and optical spectroscopy
  with AAOmega. At present the $i$-band imaging and AAOmega
  spectroscopic surveys are completed, while the other observations
  are ongoing. Table~\ref{depths} summarizes the depths and completion
  of the imaging surveys which are described in Sects.~\ref{optical}
  and \ref{VISTA}. In the table we list both the target depths and
  those measured from the data\footnote{The noise inside the aperture
    can be estimated from the flux and its uncertainty derived by
    SExtractor (see eq.~61 of SExtractor User Manual;
    www.astromatic.net/software/sextractor}. For the characteristics
  of the spectroscopic survey see Sect.~\ref{AAO}.

\begin{table}
  \centering
\caption{{\bf Assumed m$^\star$ values.}}
  \begin{tabular}{ccc}
    \hline \hline 
{\bf Band} & {\bf m$^\star$} & {\bf reference} \\
\hline
$B^a$ & 15.35 & \cite{SOSI} \\

$R^a$ & 14.52 & \cite{SOSI} \\ 

$K^a$ & 11.70 & \cite{ACCESSI} \\

$r_{AB}^b$ & 15.00 & \cite{SOSI} \\
\hline
\hline
\end{tabular}
\begin{tabular}{l}
\end{tabular}
\begin{tabular}{l}
a) Vega photometric system. \\
b) Derived from the $R$-band value once converted in $r_{AB}$ \\
with \cite{BR07} and \cite{FSI93}. \\
\end{tabular}
\label{mstar}
\end{table}

\begin{table*}
  \centering
\caption{{\bf ShaSS imaging: depths and current coverage.}}
  \begin{tabular}{cccccr}
    \hline \hline 
{\bf Band} & {\bf Exp. Time} &\multicolumn{2}{c}{{\bf depth [5$\sigma$]}\,$^a$} & {\bf seeing}\,$^b$ & {\bf complete} \\ 
& [s] & {\bf target} & {\bf measured} & [arcsec FWHM[ & [\%] \\
\hline
$u$ & 2955 & 24.5 & 24.3 & 0.8-1.1 & 48 \\

$g$ & 1400 & 24.2 & 24.8 & 0.6-1.0 & 43 \\ 

$r$ & 2664 & 24.2 & 24.3 & 0.6-0.8 & 61 \\

$i$ & 1000 & 22.4 & 23.2 & 0.5-1.0 & 100 \\

$K$ & 1620 & 20.4 & 20.3 & 0.6-1.0 & 60 \\
\hline
\hline
\end{tabular}
\begin{tabular}{l}
\end{tabular}
\begin{tabular}{l}
a) Within a 3\,arcsec diameter aperture. \\
b) Range of seeing estimated from the already observed fields. \\
\end{tabular}
\label{depths}
\end{table*}

\subsection{VST-ACCESS survey}
\label{optical}

The new optical survey (PI: P. Merluzzi),  conceived in the
  framework of the ACCESS project and named after it VST-ACCESS, is
being carried out using the Italian INAF Guaranteed Time of
Observations (GTO) with OmegaCAM at the 2.6m ESO telescope VST
\citep{SCA12} located at Cerro Paranal (Chile). The corrected field of
view of 1$^\circ$x 1$^\circ$ allows the whole ShaSS area to be covered with
23 VST fields as shown in Fig.~\ref{VST-ACCESS}. Each of the
contiguous VST-ACCESS fields is observed in four bands: $ugri$. Red
dots in Fig.~\ref{VST-ACCESS} denote the 1676 spectroscopic
supercluster member galaxies ($11300 < V_h < 17000$\,km\,s$^{-1}$)
available from literature at the time of the survey planning. The
X-ray centres are indicated by crosses for all the known clusters
except AS\,0726, whose centre is derived by a dynamical analysis.

\noindent
\underline{{\it Survey depths.}} We will achieve our scientific goals
studying the global and internal physical properties of Shapley
galaxies down to $m^\star$+6. In particular, we need to i) derive
accurate morphology as well as structural parameters
($\delta${\it log}$r_e\sim0.04$ and $\delta n_{Ser} \sim1$) and detect
some of the observational signatures related to the different
processes experienced by supercluster galaxies (e.g. extraplanar
material); ii) estimate accurate colours, photo--$z$ \citep[$\delta
z<0.03$, see][]{CEL12} and stellar masses; iii) evaluate the
star-formation rates and resolve the star forming regions at least
for the subsample of brighter galaxies. The required SNR depends on
which galaxy property is measured: it is higher for the morphological
analysis and resolving internal properties/structures
\citep[SNR$\sim$100 in our 3\,arcsec aperture, see][]{CBJ00,HMB07},
but it can be significantly lower for accurate measurements of
aperture photometry and colours (SNR$\sim$20). In the latter case,
however, the minimum required SNR should be achieved in all bands.

We chose to use mainly the $r$-band imaging for the morphological
analysis, so the $r$ band defines the survey depths in all bands. In
$r$ band $m^\star_r \sim 15$ (AB magnitude, see Table~\ref{mstar}) and
at $m^\star_r$+6=21 we require SNR=100 for the
morphological/structural analysis. The completeness magnitude of the
catalogue in $r$ band is instead defined by the star/galaxy separation
(see Sect.~\ref{DA}) which we estimate to be robust down to
$r$=23.5\,mag (SNR$\sim$10 within a 3\,arcsec aperture), corresponding
to a limiting magnitude $r$=24.2\,mag (SNR=5). We are collecting the
$r$-band imaging under the best observing conditions having a median
seeing FWHM$\sim$0.8\,arcsec corresponding to 0.75\,kpc at $z\sim
0.05$. Additionally, the $r$ imaging is fundamental to our weak
lensing analysis, to ensure a sufficient density of lensed background
galaxies with shape measurements.

The typical colours at $z\sim 0.048$ for red sequence galaxies in the
AB photometric system are: $u{-}g{\sim}1.4$, $g{-}r{\sim}0.8$,
$r{-}i{\sim}0.35$ according to stellar population models
\citep[][$\tau{=}3.0$\,Gyr, $Z=Z_\odot$]{BC03}. This approximation
allowed us to estimate the required depths for the other three bands
which all should provide complete galaxy samples down to $m^\star$+6
with SNR$\ge 20$. 

\noindent
\underline{{\it Observation strategy.}} The 1 square degree
unvignetted field of view is sampled at 0.21\,arcsec per pixel by
OmegaCAM with a 16k$\times$16k detector mosaic of 32 CCDs which constitute
the science array. The detector mosaic presents gaps up to 25\,arcsec
and 85\,arcsec wide in X and Y direction, respectively. To bridge the
gaps, we chose the dither offsetting mode with a diagonal pattern and 5
exposures for $i$, $g$ and $u$ bands. In order to reach the required
depth, cover the gaps and avoid saturation in the centre of bright
galaxies, the $r$-band images are instead obtained with 9 exposures and
smaller offsets. The contiguous pointings of VST-ACCESS are overlapped
by 3\,arcmin. This strategy allows, for each band, to use a few
pointings observed under photometric conditions in each run to
calibrate all the other fields as well as to check the photometric
accuracy.

\begin{table}
  \centering
\caption{{\bf VST-ACCESS observations P88-P91.}}
  \begin{tabular}{cll}
    \hline \hline 
{\bf Field} & {\bf Bands} & {\bf ESO Periods}$^a$ \\
\hline
F01 & $i$ & P89 \\
F02 & $i$ $r$ & P89 P91 \\
F03 & $i$ $r$ $g$ $u$ & P90 P91 P88 P88 \\
F04 & $i$ & P90 \\
F05 & $i$ & P90 \\
F06 & $i$ $r$ & P89 P91 \\
F07 & $i$ $r$ & P89 P91 \\
F08 & $i$ $r$ $g$ $u$ & P88 P89 P88 P90 \\
F09 & $i$ & P91 \\
F10 & $i$ & P90  \\
F11 & $i$ $r$ $u$ & P89 P91 P91\\
F12 & $i$ $r$ $g$ $u$ & P89 P90 P91 P91 \\
F13 & $i$ $r$ $g$ $u$ & P90 P90 P89 P88 \\
F14 & $i$ & P90 \\
F15 & $i$ & P90 \\
F16 & $i$ $r$ $g$ $u$ & P89 P90 P91 P91 \\
F17 & $i$ $r$ $g$ $u$ & P89 P90 P91 P91 \\
F18 & $i$ $r$ $g$ $u$ & P88 P90 P90 P89 \\
F19 & $i$ & P91 \\
F20 & $i$ & P91 \\
F21 & $i$ $r$ $g$ $u$ & P88 P90 P91 \\
F22 & $i$ $r$ $g$ $u$ & P88 P90 P90 \\
F23 & $i$ $r$ $g$ $u$ & P88 P90 P90 \\
\hline
\hline
\end{tabular}
\begin{tabular}{l}
\end{tabular}
\begin{tabular}{l}
a) For each band in column 2 the period of observations is indicated. \\
\end{tabular}
\label{PP}
\end{table}

\begin{figure}
\includegraphics[width=80mm]{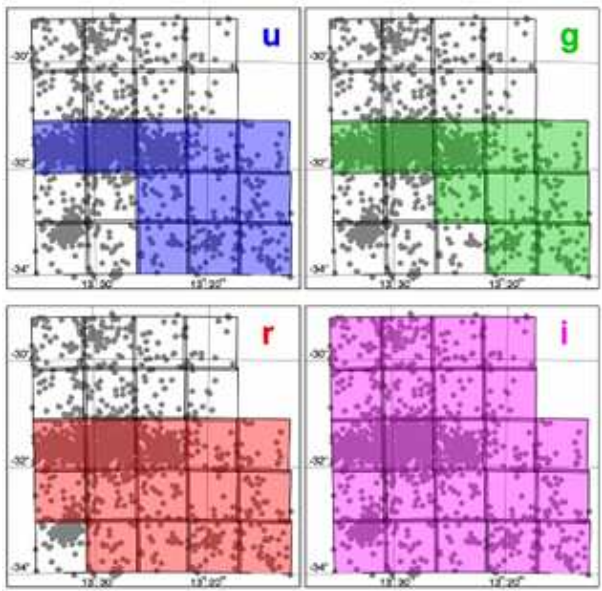} 
\caption{The coloured squares show the VST-ACCESS field coverage at the end of
  ESO-P91 (September 2013). Gray dots are the supercluster members from literature.}
\label{coverage}
\end{figure}

\noindent
\underline{{\it Survey strategy.}} To carry out the optical survey
with VST about 100 hours of telescope time are foreseen. The
observations started in February 2012 and are in progress. In the
first four ESO semesters a total of 48\,h have been allocated of which
87 per cent carried out. We show the coverage at the end of the first
two years of the survey (September 2013) in Fig.~\ref{coverage} and in
Table~\ref{PP} the distribution of the observations across this
period. The following strategy is chosen to set the priority of the
fields to be observed.
\begin{description}
\item[1)] To map the whole area in the $i$ band in order to have a
  magnitude-limited galaxy catalogue with high astrometric and
  photometric accuracy. This catalogue was mandatory to carry out the
  spectroscopic survey (see below).
\item[2)] To collect multi-band optical imaging in the region of the
  SSCC where we have already available the multi-band data-set (see
  Sect.~\ref{CD} ) which allows a cross-check of some of the
  quantities derived from the optical data, e.g. SF
  indicators.
\item[3)] To map in optical bands the southern 15\,deg$^2$ first,
  including the two Abell clusters A\,3554 and A\,3560 and probable
  filaments connecting these clusters to the SSCC and which is already
  covered by the VISTA $K$-band survey (see Sect.~\ref{VISTA}).
\item[4)] To complete then the wavelength coverage of the northern
  8\,deg$^2$ starting from the eastern side (A\,3559).
\end{description}

\begin{figure}
\includegraphics[width=84mm]{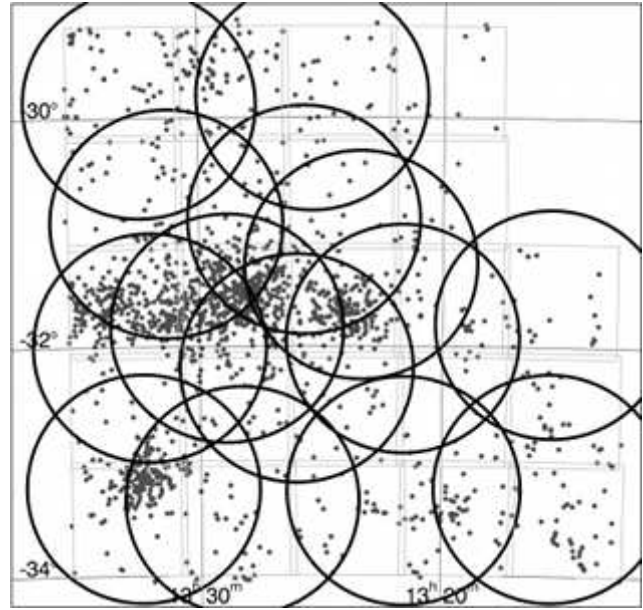} 
\caption{Overlaid to the VST-ACCESS coverage (gray boxes), the 14
  2deg-diameter AAOmega pointings (circles). Gray dots as in
  Fig.~\ref{coverage}. Two VST-ACCESS pointings (top-rigth) are not
  covered by the spectroscopic survey.}
\label{AAO_conf}
\end{figure}

\subsection{AAOmega spectroscopic survey}
\label{AAO}

A spectroscopic survey of 21\,deg$^2$ of the ShaSS area\footnote{At
  the time of the AAOmega observations two of the VST fields had not
  yet been observed.} was carried out with the AAOmega spectrograph at
the 3.9m Anglo Australian Telescope in May 2013 (PI: P. Merluzzi). The
main aim of this survey is to trace the structure of the SSC, beyond
the known galaxy clusters, including filaments and groups, in order to
describe the local environment in as much detail as possible. We also
aimed to reach high completeness in stellar mass and SFR, and for such
reason the priorities given to the targets were based on WISE W1
3.4$\mu$m (as a proxy for stellar mass) and W3 12$\mu$m (as a proxy
for SFR) magnitudes. The detailed analysis of the spectroscopic
survey will be presented in a forthcoming article. Here, we note brief
details of this process and the pertinent numbers of galaxies
recovered.

AAOmega is a dual-beam fibre-fed spectrograph, allowing the
acquisition of up to 392 simultaneous spectra of objects in a two
degree diameter field on the sky. Our observations were carried out
with gratings 580V and 385R on the blue and red arm respectively,
covering the wavelength range 370\,--\,880\,nm at a resolution of
$\lambda / \Delta \lambda\sim $ 1300. The survey consisted of 14
AAOmega pointings of 1.5\,h integration time (including several
repeats to constrain the pair-wise blunder rate\footnote{This is the
  incidence rate where one measures and re-measures (independently)
  the redshift of a target galaxy. If these two values are different
  by $>$600\,km\,s$^{-1}$, it is a pair-wise blunder \citep{C01}.}),
whose centres were determined using a simulated annealing algorithm in
the same manner as \citet{DJB10} which allows to optimize the number
of targets. Figure~\ref{AAO_conf} shows the configuration of the 14
AAOmega pointings superimposed to the VST fields.

The data were reduced in a standard manner using {\sc 2dfdr}
\citep{L02} which included a Laplacian edge detection to remove
incident cosmic rays. Redshifts where obtained using the {\sc runz}
code common to 2dFGRS, 2SLAQ, GAMA, WiggleZ \citep{C01,C06,D11,DJB10}
which involves Fourier space correlation of each spectrum to a wide
batch of template spectra.

We obtained 4037 new redshift measurements in the whole area, which,
combined with pre-existing measurements, give a total of 6130
redshifts. In the redshift range of the SSC assumed here (see
Fig.~\ref{histo}), the available redshifts are now 2281, of which 915
have been obtained with our AAOmega survey (see
Sect.\ref{dens}). Beyond the SSC, 3014 AAOmega redshifts encompass the
large-scale background structure, unveiling remarkable peaks in the
range $z$=0.07--0.25 which were mostly unknown before.

The AAOmega survey is 80 per cent complete down to $i=17.6$\,mag,
W1=14.7\,mag and W3=10.7\,mag. These two last translate into a
completeness of 80 per cent in stellar mass and SFR of respectively
$\mathcal {M}_\star\sim $8.7$\times 10^{9}$ M$_{\odot}$ and
SFR$\sim$0.7\,M$_{\odot}$\,yr$^{-1}$ at the supercluster
redshift. The value of the stellar mass is obtained by means of
  the W1-stellar mass calibration as determined by matching SDSS DR7
  galaxies at the same distance as the SSC, with stellar mass
  estimates from \citet{SRC07} and \citet{BR07}, to the WISE all
  sky-catalogue, and taking the best-fit linear relation between W1
  flux and stellar mass. The SFR is evaluated from the W3 magnitude
  using eq.~1 of \citet{DYT12}.

\begin{figure}
  \includegraphics[width=84mm]{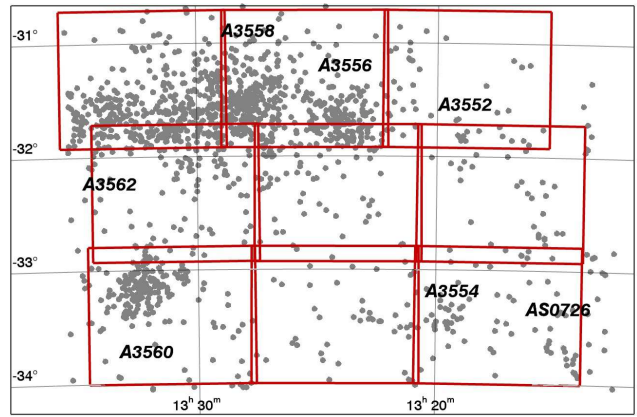}
  \caption{The nine VISTA tiles (red rectangles) mapping the southern
    SShaSS area. Gray dots as in Fig.~\ref{coverage}.}
\label{VISTA_F}
\end{figure}

\subsection{Shapley-VISTA $K$-band survey}
\label{VISTA}

The NIR survey (PI: C.~P. Haines) is being carried out with VIRCAM at
the 4m ESO telescope VISTA located at Cerro Paranal (Chile). These
$K$-band observations are accomplished using the Chilean GTO.
 
VIRCAM covers 0.59\,deg$^2$ per single pointing (paw-print) with 16
2048pxl$\times$2048pxl detectors. The gaps between the arrays amount
to 90 per cent and 42.5 per cent of the detector size along the X and
Y axis respectively. So, to obtain a contiguous coverage of the
$1.5^\circ\times 1^\circ$ field of view six offsetting paw-prints (a
tile) are needed. The mean pixel size is 0.339 arcsec. Almost all the
ShaSS area can be covered with 15 VIRCAM tiles. At present, the 9
  tiles in Fig.~\ref{VISTA_F} have been observed.

Combination of the $K$-band data with the VST optical imaging ($ugri$)
we will derive accurate stellar masses by means of stellar population
models constrained by the observed optical and infrared colours. This
will allow us to robustly measure the SMF and distinguish the
contributions of star-forming and passive galaxies to the SMF in
different mass ranges down into the dwarf regime.

\noindent
\underline{{\it VISTA survey depths.}} To achieve our scientific goals
we need to estimate stellar masses down to $\mathcal{M}=10^7$M$_\odot$
(see Sect.~\ref{ACCESS}) corresponding to $K\sim$19.6 ($\sim
K^\star$+8, see Table~\ref{depths}) which is therefore the requested
completeness limit for accurate aperture photometry (SNR$>$10 for a
point source in 3\,arcsec aperture). In order to estimate the
NIR-optical colour gradients with 20-30 per cent accuracy using
VST-ACCESS and Shapley-VISTA data, a SNR$\sim$40-50 is required and
according the survey depth this can be achieved at magnitudes brighter
than $K\sim$17.7 corresponding to $K^\star$+6 \citep[$\ge
10^{8.75}$M$_\odot$,][]{ACCESSI}.

\noindent
\underline{{\it VISTA observations and survey strategy.}} The
  13.5\,deg$^{2}$ southern ShaSS regions was observed in the $K_{s}$
  band by VISTA in April--May 2014 (093.A-0465: 18\,hours allocated),
  covered by a $3{\times}3$ mosaic of VIRCAM tiles as shown in
  Fig.~\ref{VISTA_F}. Each stacked paw-print image consists of
  $9{\times}9$\,s exposures, repeated using 5-point jitter pattern
  with a maximal offset from the central position of 30\,arcsec. Each
  tile of six paw-prints was observed a second time, such that each
  point within the tile region was covered by four stacked
  paw-prints\footnote{Each point within a tile is covered by 2 of the
    6 paw-prints.}, giving an exposure time of 1620\,s per-pixel. As
  for the VST-ACCESS survey, the tiles are slightly overlapped
  (${\sim}8$\,arcmin in Y and ${\sim}3$\,arcmin in X) to confirm the
  consistency of the photometry from one tile to another. Initial data
  reduction steps were performed at the Cambridge Astronomical Survey
  Unit (CASU) using a software pipeline developed specifically for the
  reduction of VIRCAM data, as part of the VISTA Data Flow
  System\footnote{http://casu.ast.cam.ac.uk/surveys-projects/vista/vdfs}
  \citep[VDFS][]{ILH04}. VISTA $K_{s}$ magnitudes are calibrated onto
  the Vega magnitude photometric system using unsaturated 2MASS stars
  in the image, based on their magnitudes and colours in the 2MASS
  point source catalogue. The resultant stacked $K$-band images have
  FWHMs in the range 0.6--1.0\,arcsec. The $K$-band magnitude
  detection limit at 5$\sigma$ within a 3\,arcsec aperture turns out
  to be 20.3\,mag.

We aim to complete observations of the remaining northern 8\,deg$^{2}$
region with VISTA in 2015.

\subsection{Complementary data}
\label{CD}

The Wide-field Infrared Survey Explorer \citep[WISE,][]{WISE10} is a
NASA Explorer mission that observed the entire sky in 2010 in four
near/mid-infrared bandpasses: 3.4\,$\mu$m (W1), 4.6\,$\mu$m
(W2), 12\,$\mu$m (W3) and 22\,$\mu$m (W4). All the data have been
reduced, calibrated and released to the
public\footnote{http://irsa.ipac.caltech.edu/Missions/wise.html}. The
WISE satellite made twice as many passes of the region covered by the
ShaSS survey as it did on average for most areas of the sky (the
ecliptic poles excluded), and so the limiting magnitudes are slightly
deeper than the typical WISE depths reaching W1=16.96 (Vega magnitude)
at SNR=10. This corresponds to $\sim$m$^\star$+5 for galaxies in the
SSC. The W2 filter (4.6\,$\mu$m) is a magnitude shallower. We
typically reach W2=15.26 at a SNR of 10. The W3 filter (12\,$\mu$m)
reaches a depth of W3=11.12 mag (1.0\,mJy) at a SNR of 10. This
corresponds to a SFR of 0.46\,M$_\odot$yr$^{-1}$ for galaxies in the
SSC \citep[eq.~1 of][]{DYT12}. The W4 filter (22\,$\mu$m) reaches a
depth of W4=7.58 (7.7\,mJy) at SNR of 10. This corresponds to a SFR of
$\sim$2.1\,M$_\odot$yr$^{-1}$ for galaxies in the SSC. The resolution
of the WISE bands are 6.1\,arcsec, 6.4\,arcsec and 6.5\,arcsec FWHM in
bands W1, W2 and W3. W4 instead has a FWHM of 12.0\,arcsec.

To cross-correlate WISE and optical catalogues, we use the
  software STILTS\footnote{http://www.star.bris.ac.uk/~mbt/stilts}
  searching for the closest match within a 3\,arcsec radius between
  the VST-ACCESS and WISE detections. Due to higher spatial
  resolution, the astrometry of the optical images is more accurate
  and source deblending is checked to avoid multiple detections of a
  single extended source. In the WISE catalogue, excluding the very
  extended and bright sources, i.e. few resolved nearby galaxies,
  there are very few multiple detections of extended sources. This
  allows us to associate each optical ($gri$) detection only with one
  IR detection. The few cases of multiple detection of extended
  sources in the IR images are then fixed in the final
  cross-correlated optical-IR catalogue. This approach has been
  adopted for the $i$-WISE catalogue used for the spectroscopic
  survey. We note also that nearby IR resolved galaxies were not
  spectroscopic targets, since their redshifts were already available
  from the literature.

The ShaSS data are complemented in the central 2-3\,deg$^2$ by {\it
  Spitzer}/MIPS 24/70$\mu$m photometry and {\it GALEX}
near-ultraviolet and far-ultraviolet imaging which allowed us to
produce a complete census of SF (both obscured and unobscured),
extending down to SFRs $\sim$0.02--0.05\,M$_\odot$yr$^{-1}$, i.e down
to levels comparable to the SMC. Although these data do not cover the
whole ShaSS region they are fundamental for our survey. The W3 data
will enable us to measure the SFR down to 0.2\,M$_\odot$yr$^{-1}$ at
5$\sigma$, while the W1-W3 colour allows us to reproduce and map the
bimodal galaxy distributions seen in the f$_{24}/f_K$ galaxy colours
in the supercluster core \citep{ACCESSII}, but over the entire
supercluster region, splitting the supercluster galaxies into
star-forming, transitional and passive populations. This is possible
because we verified that a strong linear correlation exists between
W1-W3 colour and f$_{24}/f_K$.

Near-infrared $K$-band imaging from the United Kingdom Infrared
Telescope with the Wide Field Infrared Camera are also available for
3\,deg$^2$ in the SSCC providing a complete galaxy sample down to
$K$=18 \citep{ACCESSI}. XMM-Newton archive data are available for
the SSCC region and for A\,3560 and will enable us to map the ICM gas and
to identify possible shock fronts due to cluster mergers and then to
investigate the effects of such events on galaxy properties.

For a subsample of supercluster galaxies, we have been obtaining
integral-field spectroscopy with WiFeS \citep{Dopita07} on the
Australian National University 2.3m telescope at Siding Spring in
Australia. We are targeting a few bright ($m<m^\star$+1.5)
supercluster galaxies showing evidence of undergoing
transformation. All these galaxies are selected by either disturbed
morphology, such as asymmetry and tails, or evidence of star-formation
knots \citep{ACCESSV}. For these objects, we also have obtained H$\alpha$
imaging with Maryland-Magellan Tunable Filter
\citep[MMTF;][]{veilleux} on the Magellan-Baade 6.5m telescope at the
Las Campanas Observatory in Chile.

\begin{table*}
\centering
\caption{{\bf Typical values for the absolute photometric calibration}}
\begin{tabular}{c c c c l}
\hline \hline
{\bf Night} & {\bf Band} & {\bf ZP}  & {\bf colour term} & {\bf extinction} \\
            2013-02-13 & u & 23.261 $\pm$ 0.028 & 0.026 $\pm$ 0.019 ($u-g$) & 0.538  \\
            2013-03-16 & g  & 24.843 $\pm$ 0.006 & 0.024 $\pm$ 0.006 ($g-i$) & 0.18  \\
            2012-04-29 & r  & 24.608 $\pm$ 0.007 & 0.045 $\pm$ 0.019 ($r-i$) & 0.1  \\
            2012-02-27 & i  & 24.089 $\pm$ 0.01 & -0.003 $\pm$ 0.008 ($g-i$) & 0.043  \\
            \hline \hline
\end{tabular}
\label{ZP}   
\end{table*}

\section{VST data reduction}
\label{DR}

The VST images have been processed using the VST--Tube imaging
pipeline \citep[][]{GCL12}, developed for the data produced by the VST. 

After applying the overscan correction and bias subtraction, we divide
by the master flat - a normalized combination of the dome and twilight
flats, in which the twilight flat is passed through a low-pass filter
first.

\begin{figure}
\includegraphics[width=84mm]{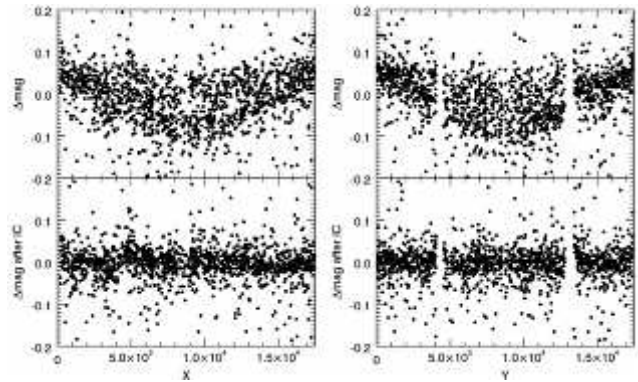}
\caption{{\it Top}: differences between the $r$-band magnitudes of
  stars in a standard field observed with VST and SDSS magnitudes
  vs. X and Y image axis. {\it Bottom}: The same $\Delta$mag after the
  IC correction has been applied.}
\label{IC}
\end{figure}

Due to differences between the electronic amplifiers, the CCDs do not
have the same gain levels. In order to have the same photometric
zeropoint (ZP) for all the mosaic chips a gain harmonization procedure
has been applied. The procedure finds the relative CCD gain
coefficients which minimizes the background level differences in
adjacent CCDs. A further correction is needed for the light scattered
by the telescope and instrumental baffling. The resulting uncontrolled
redistribution of light adds a component to the background and the
flat field will not be an accurate estimate of the spatial detector
response. Indeed, after flat-fielding, the image background will
appear flat but the photometric response will be
position-dependent. This error in the flat-fielding can be mitigated
through the determination and application of the illumination
correction (IC) map. The IC map is determined by comparing the
magnitudes of photometric standard fields with the corresponding SDSS
DR8 PSF magnitudes. The differences between the VST and SDSS
magnitudes are plotted vs. X and Y axis in Fig.~\ref{IC} before (top
panels) and after (bottom panels) the IC. Usually, the same IC map can
be used for observations carried out on the time scale of a month.

In the case of the {\it i} band it is required to correct for the
fringe pattern due to thin-film interference effects in the detector
of sky emission lines. Also this is an additive component that must be
subtracted. The fringing pattern is estimated using the
$\frac{SuperFlat}{TFlat}$ ratio where {\it SuperFlat} is obtained by
overscan and bias correcting a sigma-clipped combination of science
images and {\it TFlat} is the sky flat. To this aim, the dithering
amplitudes of the science frames used must be larger than the extended
object sizes in the same images. This is mandatory to allow the
sigma-clip procedure to efficiently remove such objects. The
difference between the above ratio across the image and its model
obtained with a surface polynomial fit is the fringing pattern. The
fringe pattern is subtracted from the image applying a scale factor
which minimizes the absolute difference between the {\it peak} and
{\it valley} values (maximum and minimum in the image background) in
the fringe corrected image.

The relative photometric calibration among the exposures contributing
to the final mosaic image is obtained comparing the magnitudes of
bright unsaturated stars in the different exposures, i.e. minimizing
the quadratic sum of differences in magnitude using SCAMP
\citep{B06}. The absolute photometric calibration is computed on the
photometric nights comparing the observed magnitude of stars in
photometric standard fields with SDSS photometry. For those fields
that are not observed in photometric but clear nights, we take
advantage of the sample of bright unsaturated stars in the overlapping
regions between clear and photometric pointings and, by using SCAMP,
each exposure of the clear fields is calibrated onto the contiguous
photometrically calibrated field. In Table~\ref{ZP} we give examples
of the fitted values for the ZP and colour term obtained using the
Photcal tool \citep{RAR04}. In several cases, the photometric standard
star fields were observed with insufficient span in airmass to do a
suitable fit. The extinction coefficient was then taken from the
extinction curve provided by ESO. The errors listed in Table~\ref{ZP}
are those of the fit. The actual errors for the ZP is given by the
r.m.s. of the detection among the different exposures of each pointing
and turns out to be less than 0.03\,mag in all bands.

The relative astrometry among the pointings is derived using SCAMP,
while for the absolute astrometry we refer to the astrometric
catalogue 2MASS obtaining $\lesssim 0.3$\,arcsec astrometric accuracy in
all bands. The image resampling, application of astrometric
solution and co-addition is performed using the tool SWARP
\citep{B02} which produces the final stacked image with a weighted
average. At the stage of the co-addition the scale factors for
relative and absolute photometric calibration are applied.

\section{Analysis of the optical images}
\label{DA}

Once the mosaic images were obtained, we proceeded with the catalogue
extraction and measured the photometric accuracy and
completeness. Details of ShaSS catalogue extraction and release will
be given in a companion article (Mercurio et al. in preparation).

 We used the software package SExtractor \citep{BA96} estimating the
 background locally and using a Gaussian filter for source
 detection. For each source we measured magnitudes in different
 apertures as well as Kron and PSF magnitudes.

To distinguish between stars and galaxies, we adopted a progressive
approach \citep{AMB13} using the following parameters provided by
SExtractor: i) the stellarity index to select point-like sources; ii)
the half-flux radius as a measure of source concentration; iii) the
new SExtractor parameter which takes into account the difference
between the model of the source and the model of the local PSF; iv)
the peak surface brightness above background; v) a final visual
inspection for objects with ambiguous values of the stellarity index.

We estimated the completeness magnitudes using both trends of the
galaxy number counts and the method by \citet{GMA99}. The catalogues
turned out to be 100 per cent complete at the total magnitudes of
23.9-24.1, 23.8-24.0, 23.3-23.5, 21.8-22.0\,mag in $ugri$ bands,
respectively. The ranges of magnitudes are due to small differences in
seeing among the VST fields.

We measured the SNR inside a 3\,arcsec aperture as function of
  magnitudes achieving SNR=20 at $ugri$ at 22.8, 23.3, 22.8,
  21.7\,mag, respectively. SNR$\sim$100 is reached at 21.1\,mag in $r$
  band (m$^\star$+6) as required. These depths enable us to study the
  galaxy population down to m$^\star$+6.

\section{Results}
\label{results}

\begin{figure}
\includegraphics[width=80mm]{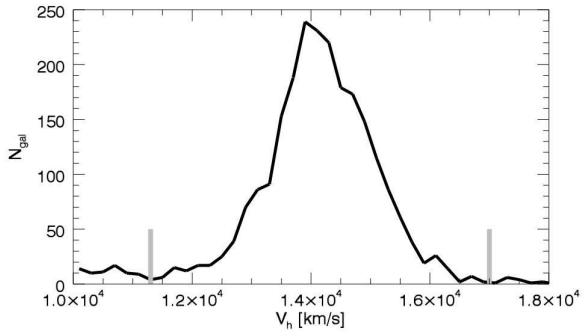} 
\caption{Redshift distribution of galaxies in the ShaSS around the
  SSC. Gray vertical lines indicate the redshift range adopted here
  for the SSC: 11300-17000\,km\,s$^{-1}$.}
\label{histo}
\end{figure}

\begin{figure*}
\includegraphics[width=160mm]{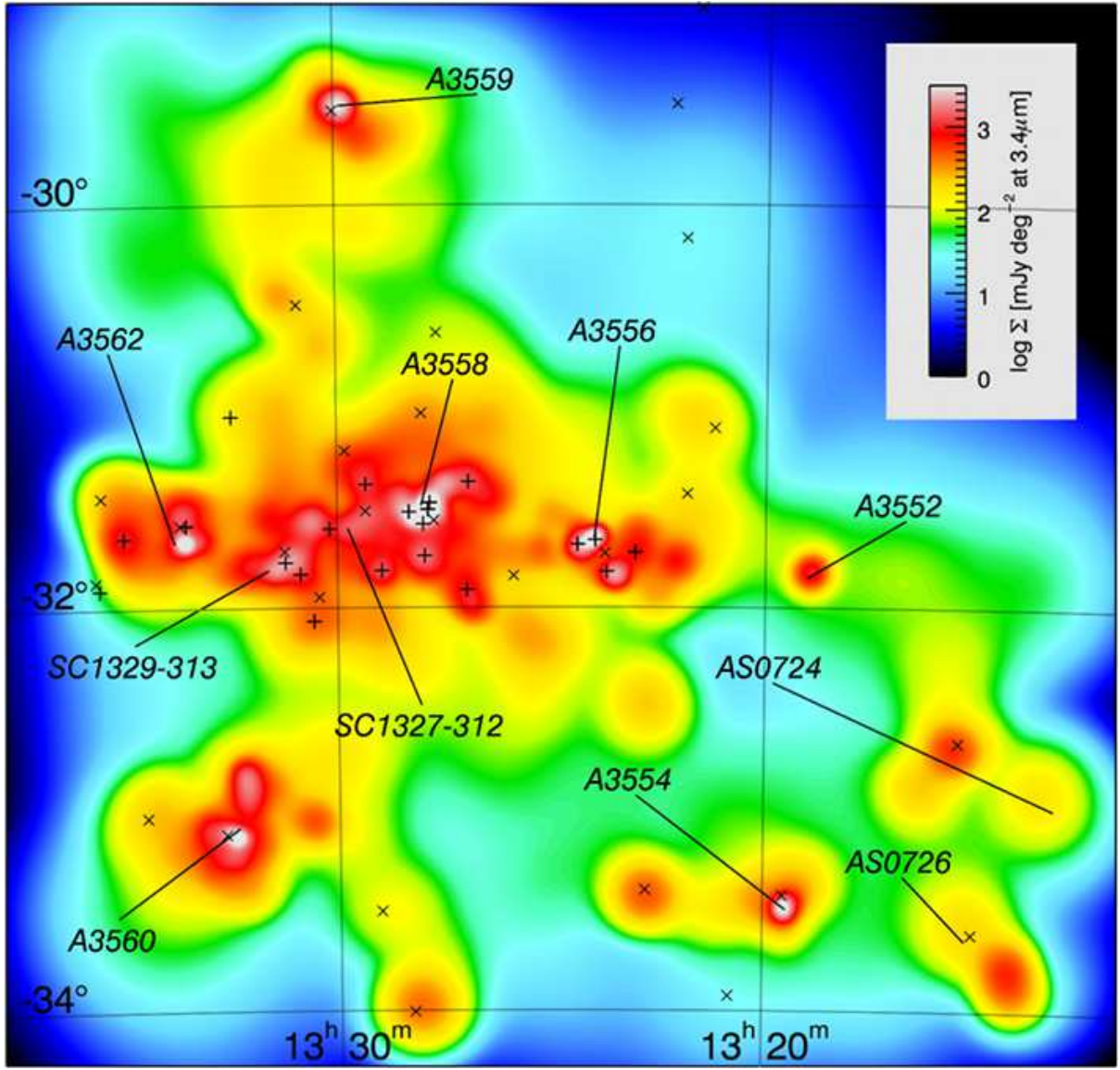} 
\caption{ShaSS density map in unit of mJy per square degree at
  3.4\,$\mu$m. Abell clusters and groups are labelled, the black
  straight lines pointing on the X-ray centre for all the systems
  except AS\,0726. Cluster substructures and groups identified by
  \citet[][$+$]{BPR98} and by \citet[][$\times$]{RMP06} are shown in
  the map. The upper right corner is not covered by ShaSS.}
\label{D_map}
\end{figure*}

In this section we derive the first quantitative characterization of
the environment in the survey area in terms of stellar mass density.
We also derive the dark matter distribution in the very centre of the
SSCC including the cluster A\,3558 and the poor cluster SC\,1327-312.

\subsection{Galaxy density}
\label{dens}

To map the structure of the supercluster, and determine its extent in
redshift space and across the plane of the sky, we take advantage of
our redshift survey which allows us to demarcate the supercluster in
redshift space as lying within the recession velocities
11300--17000\,km\,s$^{-1}$ (see Fig.~\ref{histo}). These cuts select a
supercluster sample of 2281 galaxies across the ShaSS area.

Each galaxy was weighted ($w_{j}$) according to the inverse
probability of it having been observed spectroscopically. Firstly,
each galaxy which could have been targeted for spectroscopy (${\rm
  W1}<15.0$, $i<18.0$) or was a spectroscopic member of the
supercluster, was given an initial equal weight of 1.0. For each of
these galaxies lacking a redshift, its weight was transferred equally
to its ten nearest neighbours with known redshift that also had the
same priority level in our AAOmega spectroscopic survey. This results
in galaxies without redshifts having zero weight, while galaxies in
regions where the spectroscopic survey is locally 50$\%$ complete having
weights of 2.0. The transferring of weight only within priority levels
ensures that we can account statistically for the systematic
differences in spectroscopic completeness from one level to another,
as well as mapping the local spatial variations in completeness.

Each galaxy $j$ belonging to the supercluster is represented by a
Gaussian kernel whose transverse width is iteratively set to
$\sigma_{0} [\rho_{j}(\mathbf{x}, z)/\bar{\rho} ]^{-1/2}$, where
  $\bar{\rho}$ is the geometric mean of the $\rho_{j}$, and a fixed
  radial width of 700\,km\,s$^{-1}$. Each galaxy is normalized by the
  weight parameter $w_{j}$ to account for spectroscopic
  incompleteness. The transverse kernel width for each galaxy is
  initially set to $\sigma_{0}{=}6$\,arcmin, and then iteratively
  adjusted to account for variations in the spatial density of
  galaxies, such that it typically encloses the 5 to 10 nearest
  neighbours of the galaxy, irrespective of its location within the
  supercluster.

  Figure~\ref{D_map} shows the resulting density map in which each
  galaxy is further weighted by its W1 flux as a proxy for its stellar
  mass \citep[e.g.][]{JMT13,MS14,MSV14}. The density map draws our
  attention to several features of the whole structure.

\begin{description}

\item [a)] Across the SSCC the density of supercluster galaxies is
  always higher than 315\,mJy\,deg$^{-2}$. This corresponds to
  ${\sim}7{\times}10^{12}{\rm M}_{\odot}{\rm deg}^{-2}$ based on the
  W1--stellar mass calibration obtained for the SSC galaxies (see
    end of Sect.~\ref{AAO}). Assuming a depth of 5000\,km\,s$^{-1}$
  or 71\,Mpc the stellar mass density is $9.3{\times}10^{9}{\rm
    M}_{\odot}{\rm Mpc}^{-3}$. Both the stellar mass estimates include
  a 12\% correction to account for the low-mass galaxies with
  $\mathcal{M}{<}10^{9.8}{\rm M}_{\odot}$ or W$1{>}15$ not targeted in
  our spectroscopic survey, based on the stellar mass function of
  \citet{BDL12}. This is a mean overdensity of ${\sim}40{\times}$ with
  respect to the cosmic total stellar mass density for galaxies in the
  local Universe ($z{<}0.06$) estimated by \citet{BDL12} from the GAMA
  survey \citep{D11}. We obtain a similar overdensity of
  35--40${\times}$ if we simply compare the observed W1 flux density
  of SSCC galaxies with the average W1 flux density of $W1<15$
  galaxies within the same redshift range selected from the SDSS DR7
  (8032\,deg$^{2}$). Although the SDSS is $r$-band selected, it should
  be complete to W1=14.5 at $z{=}0.048$, and only marginally
  incomplete to W1=15.0. The higher density peaks in the SSCC,
  corresponding to the cluster cores but also the groups and cluster
  sub-structures, are interconnected forming a single system. This
  implies that the galaxy distribution follows the hot gas
  distribution observed across the whole SSCC \citep[e.g.][]{KB99}.

\begin{figure}
\includegraphics[width=\hsize]{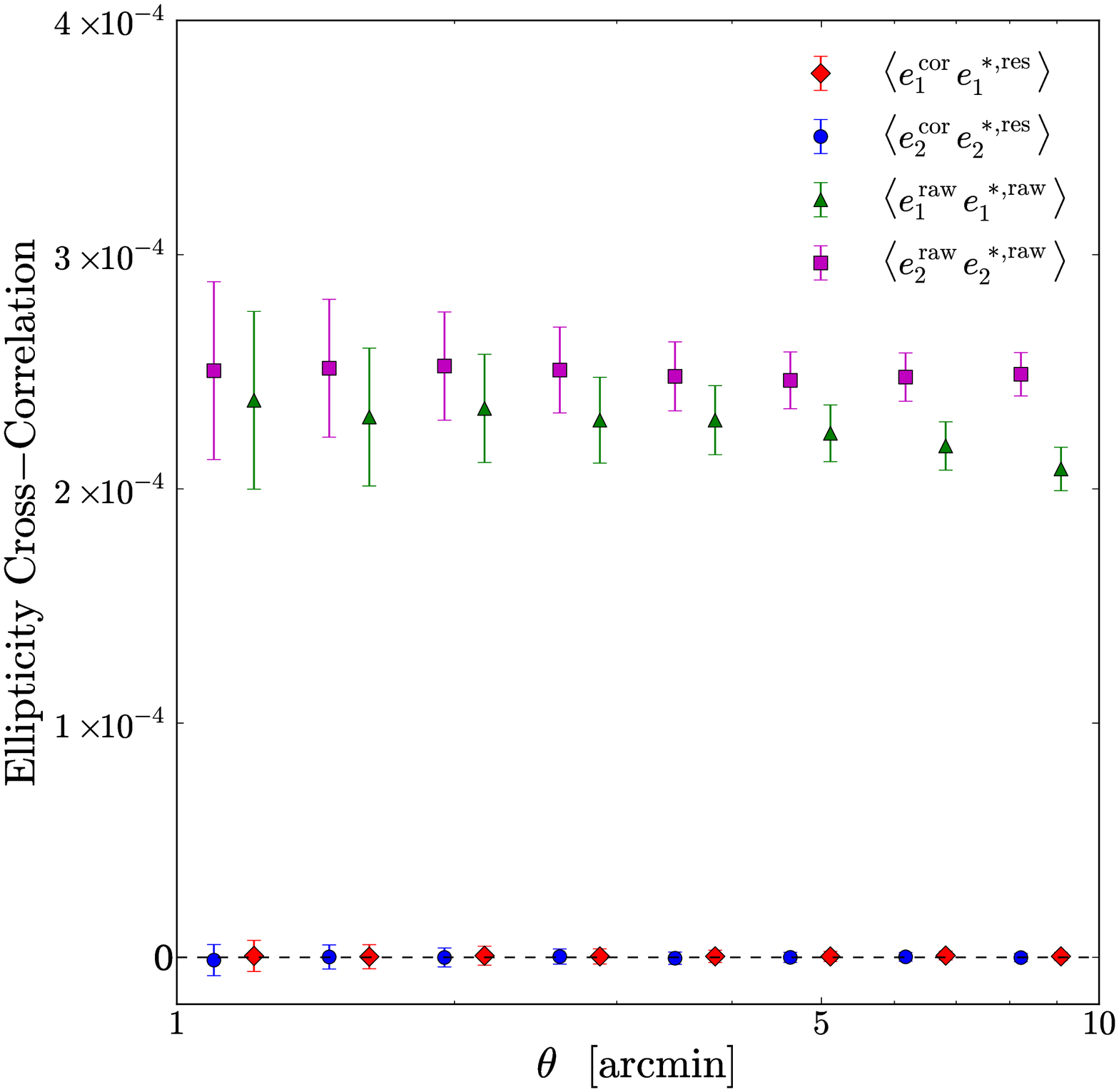}
\caption{ Cross-correlations for galaxy and stellar ellipticities,
    $\langle e_\alpha e_\alpha^{*}\rangle$. Red diamonds ($\alpha=1$)
    and blue circles ($\alpha=2$) denote the correlation function for
    corrected galaxy ellipticities, $e_\alpha^{\rm cor}$ and residual
    stellar ellipticities, $e_\alpha^{*,res}$ after correcting the PSF
    anisotropy. Green triangles ($\alpha=1$) and magenta squares
    ($\alpha=2$) are the correlation function for raw galaxy
    ellipticities, $e_\alpha^{\rm raw}$ and raw stellar ellipticities,
    $e_\alpha^{*,raw}$ before the correction. The separation radii
    are offset by $\pm5\%$.}
\label{fig:xi_ge}
\end{figure}

\item [(b)] There is clear evidence of a filament ($\sim$7\,Mpc in
  projection) heading north from the SSCC, connecting it with A\,3559,
  with W1 flux densities of ${\ga}1$50\,mJy\,deg$^{-2}$
  (${\ga}20{\times}$ overdensities). \citet{QCR00} qualitatively
  indicated a `broad arm running north' from the SSCC, but in their
  description this feature should point to A\,3557a which is located
  NW of the ShaSS region. In this direction we do not detect any
  clear overdensity.

\item[c)] Several overdensities are detected across the density
  map. Some of them can be associated to cluster dynamical
  substructures already identified by \citet[][see
  Fig.~\ref{D_map}]{BPR98}, but other density peaks are also detected,
  i.e. that W of A\,3556. The positions of the system centres as
  defined by X-ray peak and galaxy overdensity appear slightly
  different, as expected in dynamically active systems. The highest
  discrepancy between these determinations is found for AS\,0724,
  where the X-ray centre is located 35\,arcmin from the galaxy
  overdensity centre. Notice that AS\,0726 was not detected by ROSAT
  and its centre was derived by a previous spectroscopic survey
  \citep{PQC06}, likely affected by incompleteness. Our newly detected
  overdensity is actually located 16\,arcmin from the previous centre.

\item [d)] In A\,3560 two main density clumps are sharply defined
  together with a western clump. The northern clump can be related to
  the X-ray emission, elongated towards A\,3558, detected in the very
  centre of the cluster which suggested a minor merger scenario
  between the main cluster and a group located $\sim 8$\,arcmin N of
  the cluster centre \citep{VRB13}. Also the galaxy distribution shows
  an asymmetric shape elongated in the N direction, but extending a
  factor two in projection with respect to the X-ray emission
  \citep[c.f. Fig.~5 of][]{VRB13}, taking into account only the two
  denser clumps in our map. Thus the group involved in the claimed
  merger could be significantly more massive than that previously
  identified in the APM catalogue.
\end{description}

In a companion paper, we will assign galaxies to
supercluster structures/overdensities. This will be achieved through
the dynamical analysis enabling us to detect and measure the amount of
cluster dynamical substructures, and to identify possible pre-merging
clumps or merger remnants. In particular, we will i) select cluster
members \citep[e.g.][]{FGG96,GFG96}; ii) derive the centre of the most
significant peaks of each identified system applying the 2D adaptive
kernel technique to galaxy positions; iii) detect possible subclumps
and assign objects to groups \citep{ABZ94}; iv) identify the 3D
substructures combining velocity and position information
\citep{DS88}.

\begin{figure*}
\includegraphics[width=\hsize]{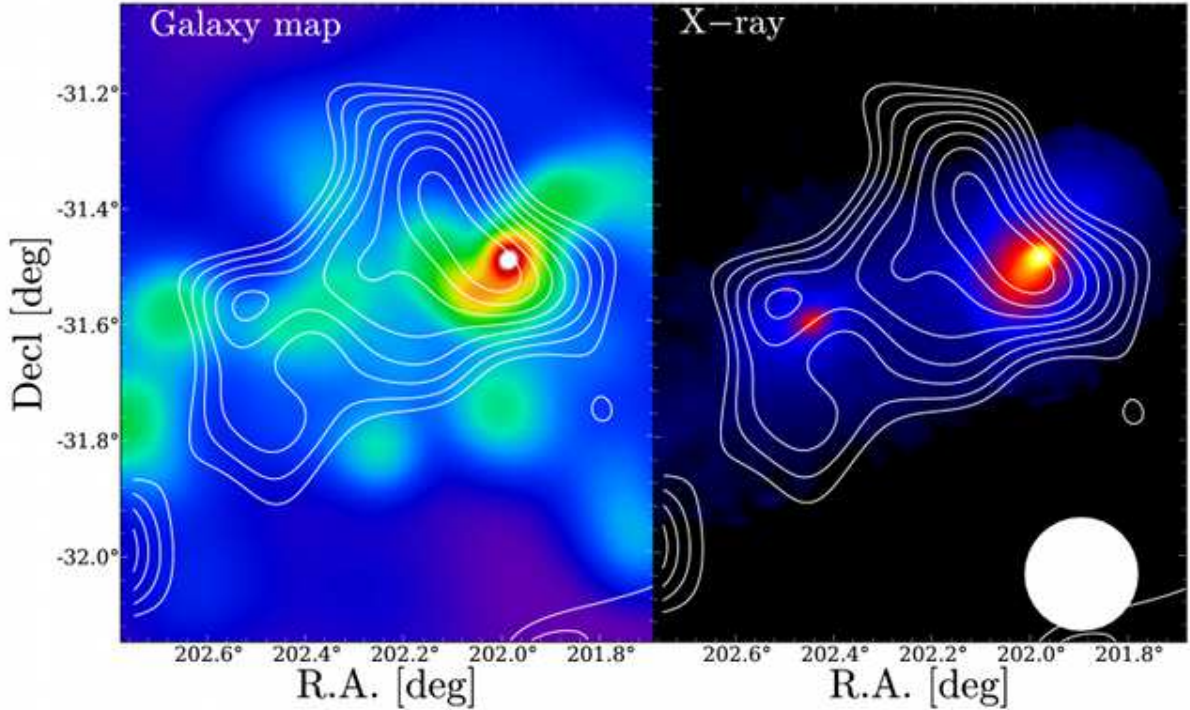}
\caption{Weak-lensing mass map of the field 8 of VST-ACCESS overlaid
  on the stellar mass density map (left) and the X-ray surface
  brightness (right). The contours of the lensing $\kappa$-field are
  in units of $0.5\sigma$ reconstruction error $\delta
  \kappa\simeq0.010$, above $1\sigma$. The resolution of the WL mass
  map is FWHM$=11.7$\,arcmin, as shown in the white circle of the
  bottom-right. The X-ray surface brightness is derived from a
  curvelet analysis of XMM-Newton images extracted in the 0.5 − 2.5keV
  energy band, that have been corrected for spatially variable
  effective exposure and background components.  The projected mass
  distribution is elongated along the large-scale X-ray filamentary
  structure. A main peak of the map is associated with the
  over-density region of member galaxies and diffuse X-ray emission of
  A\,3558. A possible secondary peak is associated with diffuse X-ray
  emission of SC\,1327-312 in the east.}
\label{fig:wlmap}
\end{figure*}

\subsection{Weak lensing mass distribution of A\,3558}
\label{WL}

We conducted a weak lensing analysis of the VST-ACCESS field 8 (see
Fig.~\ref{VST-ACCESS}) following \citet{KSB} as modified by
\citet{Okabe13,Okabe14}. We measure image ellipticity of objects
detected in the $r$-band data, $e_\alpha$, from the weighted
quadrupole moments of the surface brightness. We then correct a PSF
anisotropy of galaxy ellipticities by the function of second-order
bi-polynomials of the stellar anisotropy kernel. A cross
  correlation function of residual stellar ellipticities and the
  corrected galaxy ellipticities does not show overcorrection or
  insufficient correction as shown in Fig.~\ref{fig:xi_ge}.  Then, a
reduced distortion signal, $g_\alpha$, is estimated with a correction
of isotropic smearing effect. We select background galaxies following
\citet{Okabe13} and adopt a new method for minimizing the
contamination of member galaxies extending to the colour-colour
plane. We measure two colours, $g-i$ and $g-r$, and fit the red
sequence with a linear function. A colour offset is defined for each
galaxy by $\Delta
C\equiv\left(\Delta(g-i)^2+\Delta(g-r)^2\right)^{1/2}$, where
$\Delta(g-i)\equiv(g-i)-(g-i)_{\rm ES0}$ and
$\Delta(g-r)\equiv(g-r)-(g-r)_{\rm ES0}$ and `ES0' denotes the red
sequence galaxies. We select background galaxies by $\Delta C>0.53$
which is the lower limit to allow $1\%$ contamination level. The
number density of background source galaxies is $2.2\,{\rm
  arcmin}^{-2}$, and their mean redshift is $\langle
z_s\rangle\simeq0.49$.

The mass map is reconstructed as described in \citet{Okabe08}. The
reduced shear is pixelized into a regular grid with a Gaussian
smoothing, $G(\theta)\propto \exp[-\theta^2/\theta_g^2]$. The FWHM
resolution of the map is 11.7\,arcmin. The smoothed shear pattern is
estimated with a Gaussian kernel, $G(\theta)$, and a statistical
weight $w_{g,i}=(\sigma_{g,i}^2+\alpha^2)^{-1}$ for the $i$th galaxy,
where $\sigma_g$ is the rms error of the shear estimate and the
softening constant variance. We choose $\alpha=0.4$ as a typical value
of the mean rms of $\sigma_g$. We next invert the smoothed shear field
with the kernel \citep{Kaiser93} in Fourier space to obtain the
projected mass distribution. The resulting mass map, $\kappa$, for
field 08 of VST-ACCESS is shown in Fig.~\ref{fig:wlmap}. The mass map
is elongated along the east-west direction which is parallel to the
large-scale filamentary structure. A main peak is associated with
distributions of member galaxies and diffuse X-ray emission of
A\,3558. Another possible clump, located E of A\,3558, is likely
associated with diffuse X-ray emission of the galaxy group
SC\,1327-3136.

\begin{figure}
\includegraphics[width=\hsize]{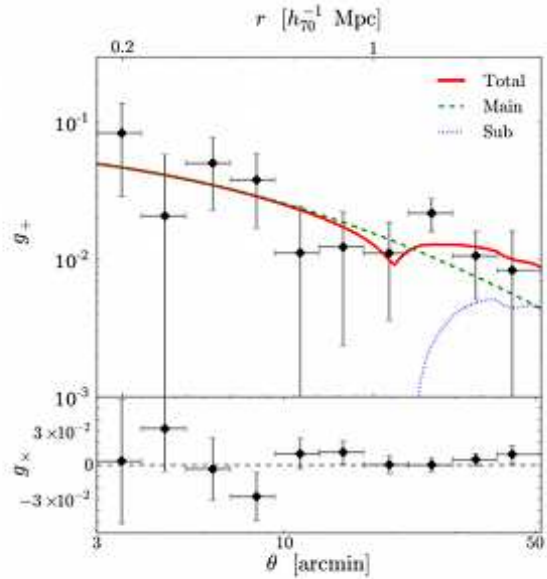}
\caption{{\it Top panel:} The tangential distortion component, $g_+$,
  as a function of the projected cluster-centric radius from the BCG, is
  estimated by azimuthally averaging the measured galaxy
  ellipticities. A bump in lensing signals is found around $r>20$.
  The profile is well described by two NFW components of A\,3558 and the
  east clump. The red solid, green dashed and blue dotted lines are
  the best-fit NFW profile for the total mass,  A\,3558 and the east
  clump, respectively. {\it Bottom panel:} The $45$ degree rotated
  component, $g_\times$, is consistent with a null signal.}
\label{gt_profile}
\end{figure}

We compute the tangential distortion component, $g_+$, with respect to
the projected cluster-centric radius from the BCG of A\,3558, shown in
Fig.~\ref{gt_profile}. The lensing signal $g+$ is decreasing as the
radius, $r_{cl}$ increases, and becomes flat or increases at
$r_{cl}>20$\,arcmin. To understand this feature, we fit the
two-dimensional shear pattern with two mass components: A\,3558 and
the east clump \citep[e.g.][]{Okabe11}. We use the Navarro Frenk \&
White (NFW) profile \citep{NFW96,NFW97} as the mass model represented
by two parameters: $M_\Delta$ which is the enclosed mass within a
sphere of radius $r_\Delta$, and the halo concentration
$c_\Delta=r_\Delta/r_s$. Here, $r_\Delta$ is the radius inside of
which the mean density is $\Delta$ times the critical mass density,
$\rho_{\rm cr}(z)$, at the redshift $z$. Since it is difficult to
constrain the concentration for the east clump, we assume a
mass-concentration relation \citep{Bhattacharyaetal2011}. The centre
of A\,3558 is fixed at the BCG position, while the position of the
secondary mass component is treated as a free parameter. The model is
described by five parameters in total.  The best-fit M$_{500}$ for
A\,3558 is $7.63_{-3.40}^{+3.88}\times10^{14}$M$_\odot$.  The total
lensing signal of the two components (solid line in
Fig.~\ref{gt_profile}) well describes the observed lensing
signals. Since A\,3558 and the east clump are embedded in the SSCC,
surrounding data are essential to further constrain the mass.

Although this is a preliminary result, the derived dark matter mass
can be compared with that evaluated from the XMM-Newton data-set. To
this aim, we computed the Y$_X$ mass proxy, defined as the product of
gas mass M$_{gas,500}$ and average temperature $kT$ \citep{KVN06}. As
described in \citet{BM08}, we inverted a gas mass profile from the
radially average surface brightness of A\,3558, then iterated about
the Y$_X$-M$_{500}$ scaling relation calibrated from hydrostatic mass
estimates in a nearby sample of clusters observed with XMM-Newton
\citep{APP10}. Note that the surface brightness of A\,3558 has been
evaluated within an angular sector excluding the eastern filament
connecting A\,3558 to SC\,1327-312.  This yielded estimates of gas
mass M$_{gas,500}=0.62\pm 0.01\times10^{14}$M$_\odot$, average
temperature $kT=4.91\pm 0.14$\,keV, and total mass M$_{500}=(4.62\pm
0.24)\times10^{14}$M$_\odot$ within r$_{500}=1160\pm 20$\,kpc, which
is consistent with the WL estimate. Previous X-ray mass estimates
based on ROSAT data gave for M$_{500}$: 8.7$\times 10^{14}$M$_\odot$
\citep{EFW97}; 6.09$\times 10^{14}$M$_\odot$ \citep{RQC00}; 8.4$\times
14^{14}$M$_\odot$ \citep[30\% uncertainty,][]{ML08}. The dynamical
mass derived by \citet{BZZ98} of A\,3558 and listed in
Table~\ref{clusters} turns out to be a factor 1.7 higher than the WL
mass, but also a factor 3 higher than X-ray mass determination
previously available as the authors stated in their
work. \citet{RMP06} computed the dynamical mass inside the virial
radius obtaining M$_{vir}$=6.7$\times 10^{14}$M$_\odot$.

\section{Summary and conclusions}
\label{sum}

The Shapley Supercluster Survey (ShaSS) aims to assess the role of
cluster-scale mass assembly on galaxy evolution searching for possible
connections between the properties of the cosmological structures
(density, dynamical status, hot-gas content, dark and luminous matter
distribution) and those of the associated galaxies (morphology,
internal structure, star formation, nuclear activity). This requires
that we extend the investigation from the cluster cores to their
outskirts, to the infalling galaxies and groups along the filaments in
a dynamically-bound network. The centre ($\sim$3\,deg radius) of the
Shapley supercluster at $z\sim 0.05$ is the optimal target to
undertake such a study enclosing a massive and dynamically active
structure showing signs of cluster-cluster mergers, enhancing the
probability to observe evidence of environmental effects on galaxy
evolution, but also providing an extraordinary variety of environments
concentrated in a small survey volume.

ShaSS includes nine Abell clusters (A\,3552, A\,3554, A\,3556,
A\,3558, A\,3559, A\,3560, A\,3562, AS\,0724, AS\,0726) and two groups
(SC\,1327-312, SC\,1329-313) covering a region of
$\sim$260\,Mpc$^2$. The survey includes the following
data-sets.

\begin{description}
\item [-] Optical ($ugri$) imaging acquired with the VLT Survey
  Telescope (PI P. Merluzzi) provides a galaxy catalogue complete
  down to $r = 23.3$ (SNR$\sim$10, limit for star/galaxy separation as
  measured from the collected data) corresponding to
  $\sim$m$^\star$+8.3 at the supercluster redshift. The multi--band
  $gri$ catalogue is complete to $\sim$m$^\star$+7.1 and the
  $u$-band catalogue to $\sim$m$^\star$+6.7. The achieved SNRs allow
  to study the galaxy population down to $\sim$m$^\star$+6 and derive
  morphological parameters (CAS+MG, see Sect.~\ref{DA}) in $r$ band to
  the same depth. The VST-ACCESS survey is ongoing.

\item [-] Near-infrared ($K$) imaging acquired with the Visible and
  Infrared Survey Telescope for Astronomy (PI C.~P. Haines) reaching
  the depth of $K\sim 19.6$ ($\sim$m$^\star$+7.9 at the supercluster
  redshift) allows to study the galaxy population down to the
  magnitude limit of the optical catalogue. The survey started in
  April 2014.

\item [-] The spectroscopic survey with AAOmega at the
  Anglo-Australian Telescope (PI P. Merluzzi) collected 4037 new
  redshifts across 21\,deg$^2$ of ShaSS. Together with the already
  available redshifts the spectroscopic sample is now 80\% complete
  down to $r\sim 18$ ($\sim$m$^\star$+3 at the supercluster redshift).

\item [-] The above dedicated surveys are complemented by
  near/mid-infrared data from the Wide-field Infrared Survey Explorer
  (WISE) in four bands (W1-W4) 3.4, 4.6, 12 and 22\,$\mu$m. Over the
  ShaSS area, we reach W1=16.96 (Vega magnitude) and W3=11.12 mag
  (1.0\,mJy) at SNR=10 corresponding to $\sim$m$^\star$+5 and SFR of
  0.46\,M$_\odot$yr$^{-1}$, respectively, for galaxies in the
  supercluster.

\end{description}

For the central 2-3\,deg$^2$, XMM-Newton, {\it Spitzer}/MIPS 24$\mu$m
and 70$\mu$m, {\it GALEX} near-ultraviolet and far-ultraviolet data
are also available, as well as targeted observations of single
galaxies providing H$\alpha$ imaging with the Maryland-Magellan Tunable
Filter on the Magellan-Baade 6.5m telescope at the Las Campanas
Observatory in Chile and integral-field spectroscopy with WiFeS on the
Australian National University 2.3m telescope at Siding Spring in
Australia.

In this first article we derived the stellar mass density distribution
based on supercluster members weighted by the W1 flux. This first
quantitative characterization of the environment for the whole region
covered by ShaSS shows a clumpy structure both in the SSCC and the
surrounding clusters with several substructures, most of them already
identified in previous works. All the clusters in the ShaSS area are
embedded in a common network. This was suggested, but only for the
SSCC where supercluster galaxies continuously populate the core
following the gas filaments connecting the Abell clusters as mapped by
X-ray observations \citep{KB99}. We estimate the mean overdensity
across the SSCC being ${\sim}40{\times}$ with respect to the cosmic
total stellar mass density for galaxies in the local Universe
($z{<}0.06$).

Some new substructures with respect to previous works have been
identified in the ShaSS density map such as those associated with
A\,3560 N from the cluster centre in the direction of A\,3558,
towards which also the X-ray emission is elongated \citep{VRB13}. The
most important new feature is however the filament connecting the SSCC
and the cluster A\,3559 as well as the less pronounced overdensity
extending from the SSCC towards A\,3560.

The other environment indicator analyzed here is the dark matter
distribution derived from the weak-lensing analysis of VST
imaging. Using this approach, we studied the central 1\,deg$^{2}$
field including A\,3558 and SC\,1327-312. The derived WL map shows
that the dark matter is concentrated in two peaks which correspond to
the rich cluster and the group, although the centres seem slightly
offset with respect to the X-ray emission and the galaxy density.
This can be due to the lower resolution of the WL map and/or ascribed
to the complex dynamical state of the SSCC. The estimated mass of
A\,3558 is $M_{500}=7.63_{-3.40}^{+3.88}\times10^{14}M_\odot$, 
consistent with the X-ray estimate of M$_{500}=(4.16\pm
0.19)\times10^{14}$M$_\odot$. We notice that the WL mass determination
will be improved extending the analysis to a larger region of the
supercluster as planned in the ShaSS project.

We conclude pointing out that the VST imaging is the first CCD
photometry covering homogeneously and continuously such a large
portion of the Shapley supercluster, with this depth and resolution
(corresponding to 0.75\,kpc at the supercluster redshift). Taking
advantage of the $i$-band and WISE W1 photometry, the AAOmega survey has
been designed and carried out to obtain a magnitude-limited redshift
sample, which was never achieved before
\citep[e.g.][]{QCR00,DPP04,PQC06}. With these characteristics, ShaSS
will build up the first multi-band homogeneous data-set of a vast
region of the Shapley supercluster and provide a fundamental local
counterpart to the supercluster surveys at higher redshifts.

\section*{Acknowledgments}
This work was conceived in the framework of the collaboration of the
FP7-PEOPLE-IRSES-2008 project ACCESS.  Based on data collected with i)
the ESO - VLT Survey Telescope with OmegaCAM (ESO Programmes
088.A-4008, 089.A-0095, 090.A-0094, 091.A-0050) and the ESO - Visible
and Infrared Survey Telescope for Astronomy with VIRCAM (ESO Programme
093A-0465) at the European Southern Observatory, Chile and ii)
Anglo-Australian Telescope and 2dF+AAOmega at the Australian
Astronomical Observatory, Australia (OPTICON proposal 2013A/014). The
optical imaging is collected at the VLT Survey Telescope using the
Italian INAF Guaranteed Time Observations. The research leading to
these results has received funding from the European Community's
Seventh Framework Programme (FP7/2007-13) under grant agreement number
312430 (OPTICON; PI: P. Merluzzi) and PRIN-INAF 2011 `Galaxy evolution
with the VLT Surveys Telescope (VST)' (PI A. Grado). CPH was funded by
CONICYT Anillo project ACT-1122. N.Okabe is supported by a
Grant-in-Aid from the Ministry of Education, Culture, Sports, Science,
and Technology of Japan (26800097) and by World Premier International
Research Center Initiative (WPI Initiative), MEXT, Japan. PM and GB
would like to thank A.~M. Hopkins for his support during the
spectroscopic observations at the Australian Astronomical Observatory
and the Universidad de Chile for the hospitality and support during
their staying. PM thanks M. Petr-Gotzens for her support in the VST
observations. The authors thank Prof. L. Campusano for his helpful
comments to the manuscript. This publication makes use of data
products from the Wide-field Infrared Survey Explorer, which is a
joint project of the University of California, Los Angeles, and the
Jet Propulsion Laboratory/California Institute of Technology, funded
by the National Aeronautics and Space Administration.  The author
thank the anonymous referee for her/his constructive comments and
suggestions.

\bibliographystyle{mn2e}

\end{document}